\documentclass[fleqn,usenatbib]{mnras}

\usepackage{newtxtext,newtxmath}

\usepackage[T1]{fontenc}

\DeclareRobustCommand{\VAN}[3]{#2}
\let\VANthebibliography\thebibliography
\def\thebibliography{\DeclareRobustCommand{\VAN}[3]{##3}\VANthebibliography}

\usepackage{lipsum}  

\usepackage{graphicx}	
\usepackage{amsmath}	

\newcommand{\vek}[1]{\mathbf{#1}}
\newcommand{\pdv}[2]{\frac{\partial #1}{\partial #2}}

\newcommand{\rot}{\mathbf{\nabla}\times}

\newcommand{\vgrad}[1]{\mathbf{\nabla}{#1}}

\newcommand*\diff{\mathop{}\!\mathrm{d}}

\DeclareMathAlphabet\mathbfcal{OMS}{cmsy}{b}{n}

\newcommand{\Bpo}{\mathbf{B_{p}}}

\newcommand{\vEMF}{\boldsymbol{{\mathcal{E}}}}
\newcommand{\mean}[1]{\left\langle #1 \right\rangle}

\newcommand{\vmean}[1]{\left\langle \mathbf{#1} \right\rangle}

\newcommand{\de}{\delta}

\title[Large-scale Magnetorotational Dynamo]{Magnetorotational dynamo can generate large-scale vertical magnetic fields in 3D GRMHD simulations of accreting black holes}

\author[J. Jacquemin-Ide et al.]{
Jonatan Jacquemin-Ide,$^{1}$\thanks{E-mail: jonatan.jacqueminide@northwestern.edu}
François Rincon, $^{2}$
Alexander Tchekhovskoy,$^{1}$
and Matthew Liska$^{3,4}$
\\
$^{1}$Center for Interdisciplinary Exploration $\&$ Research in Astrophysics (CIERA), Physics and Astronomy, Northwestern University, Evanston, IL 60202, USA\\
$^{2}$Institut de Recherche en Astrophysique et Planétologie (IRAP), Université de Toulouse, CNRS, UPS, Toulouse, France\\
$^{3}$Institute for Theory and Computation, Harvard University, 60 Garden Street, Cambridge, MA 02138, USA\\
$^{4}$Center for Relativistic Astrophysics, Georgia Institute of Technology, Howey Physics Bldg, 837 State St NW, Atlanta, GA 30332, USA\\
}

\date{Accepted XXX. Received YYY; in original form ZZZ}

\pubyear{2015}

\begin{document}
\label{firstpage}
\pagerange{\pageref{firstpage}--\pageref{lastpage}}
\maketitle

\begin{abstract}
Jetted astrophysical phenomena with black hole (BH) engines, including binary mergers, jetted tidal disruption events, and X-ray binaries, require a large-scale vertical magnetic field for efficient jet formation. However, a dynamo mechanism that could generate these crucial large-scale magnetic fields has not been identified and characterized. We have employed 3D global general relativistic magnetohydrodynamical (MHD) simulations of accretion disks to quantify, for the first time, a dynamo mechanism that generates large-scale magnetic fields. This dynamo mechanism primarily arises from the nonlinear evolution of the magnetorotational instability (MRI). In this mechanism, large non-axisymmetric MRI-amplified shearing wave modes, mediated by the axisymmetric azimuthal magnetic field, generate and sustain the large-scale vertical magnetic field through their nonlinear interactions.
We identify the advection of magnetic loops as a crucial feature, transporting the large-scale vertical magnetic field from the outer regions to the inner regions of the accretion disk. This leads to a larger characteristic size of the, now advected, magnetic field when compared to the local disk height. We characterize the complete dynamo mechanism with two timescales: one for the local magnetic field generation, $t_{\rm g}$, and one for the large-scale scale advection, $t_{\rm adv}$. 
Whereas the dynamo we describe is nonlinear, we explore the potential of linear mean field models to replicate its core features.
Our findings indicate that traditional $\alpha$-dynamo models, often computed in stratified shearing box simulations, are inadequate and that the effective large-scale dynamics is better described by the shear current effects or stochastic $\alpha$-dynamos. 
\end{abstract}



\section{Introduction}\label{sec:intro}

\subsection{Compact mergers, disruptions and binaries}
The first direct detection of gravitational waves from a binary neutron star (BNS) merger occurred in 2017 with GW170817. This remarkable event was accompanied by the detection of a gamma-ray burst (GRB) and a kilonova, marking a significant milestone in multimessenger observations \citep[for reviews, see][]{nakar_electromagnetic_2020,margutti_first_2021}. Black hole neutron star (BHNS) mergers also hold promise for multimessenger astronomy as they are expected to exhibit similar electromagnetic emission components as BNS mergers, including kilonovae and jet emissions \citep{Paczynski1991,Mochkovitch1993,Janka1999,Rosswog2005,Metzger2010,Fernandez2015,Foucart2015,Gompertz2023}.

GRBs are often associated with jet launching from the central engine \citep{rezzolla_missing_2011,ruiz_binary_2016}, be it a neutron star or a black hole, during BHNS or BNS mergers \citep[see ][for a recent review]{metzger_kilonovae_2019}. The formation of a relativistic jet during a BHNS merger, contingent on the presence of an accretion disk, depends on various system characteristics such as mass ratio, neutron star radius, and orbital dynamics. However, accretion disks are anticipated to form post-merger for specific mass ratios and non-excessively compact neutron stars \citep{Shibata2006,Shibata2007,Foucart2011,Foucart2012a,Hayashi2021,Biscoveanu2023}.

If the jet engine is a black hole, it likely operates through the \cite{blandford_electromagnetic_1977} (BZ) mechanism. This mechanism, in turn, necessitates the presence of a large-scale vertical magnetic field threading the accretion disk. Magnetic fields also play a significant role in driving accretion and mass ejection within post-merger disks \citep{siegel_three-dimensional_2017,fernandez_long-term_2019,christie_role_2019}. Large-scale vertical fields are responsible for driving magnetohydrodynamic (MHD) outflows that effectively remove angular momentum and mass \citep{blandford_hydromagnetic_1982, ferreira_magnetized_1995}, while both small and large-scale magnetic fields facilitate angular momentum transport by generating MHD turbulence through the magnetorotational instability \citep[MRI,][] {balbus_powerful_1991}.  Consequently, understanding the emergence, regeneration, and evolution of magnetic fields is essential for understanding and constraining the dynamics of mergers.

In the context of BNS mergers, it is highly probable that magnetic fields undergo significant amplification during the collision phase. This violent event likely establishes the initial magnetic field configuration of the resulting accretion disk. The underlying mechanisms for this magnetic amplification are thought to primarily involve the Kelvin-Helmholtz instability (KHI) and magnetic winding processes \citep{price_producing_2006,kiuchi_efficient_2015,aguilera-miret_turbulent_2020,aguilera-miret_universality_2022,aguilera-miret_role_2023}. Increasingly, it is becoming clear that during the initial stages of the merger evolution ($t<100\,\,\rm{ms}$), a dynamically important toroidal magnetic field perpendicular to the axis of rotation, is generated.  This phenomenon is observed not only in simulations of BNS mergers but also in those of BHNS mergers after the initial disruption phase \citep{ruiz_jet_2018,most_accretion_2021}. However, current simulations do not completely capture the magnetic field amplification dynamics \citep{kiuchi_global_2018} and often rely on arbitrarily large magnetic fields, subgrid models or prescription for the magnetic field amplification \citep{hayashi_general-relativistic_2023,gottlieb_large-scale_2023,aguilera-miret_role_2023,most_flares_2023}.

The bulk of research efforts on magnetic field evolution in mergers has been concentrated on this initial phase. While the generated toroidal magnetic field can be exceptionally strong, it is incapable of launching jets. In this study, we aim to investigate the transformation of an initially toroidal magnetic field into a large-scale vertical magnetic field, that can launch jets. 

The question of magnetic field generation is also relevant to other astrophysical sources with less constrained initial magnetic topologies. Jetted tidal disruption events (TDEs) share similarities with BHNS mergers but occur on much larger spatial and temporal scales. These phenomena have received considerable attention, with Swift J1644+57 serving as a prominent example \citep{bloom_possible_2011, burrows_relativistic_2011}. \cite{tchekhovskoy_swift_2014} suggested that the broad-spectrum emission, spanning radio, X-rays, and $\gamma$-rays, observed in Swift J1644+57, could be explained by a dynamically significant magnetic flux anchored at the black hole event horizon, resulting in remarkably efficient jet formation. However, the main challenge arises from the limited magnetic flux available on a stellar object to facilitate such highly efficient jet launching \citep{giannios_radio_2011,kelley_tidal_2014}. Therefore, finding the mechanism to amplify the magnetic field is imperative. During the tidal disruption process, the magnetic field of the disrupted star can undergo significant amplification due to the strong shear, similar to BHNS mergers. However, absent large-scale vertical magnetic field generation, the resulting magnetic field configuration remains primarily radial or toroidal, and hence inadequate for jet launching \citep{bonnerot_magnetic_2017}. This underscores the critical importance of generating large-scale vertical magnetic fields in TDEs to facilitate powerful jet formation to account jetted TDEs like Swift J1644+57.


{ X-ray binaries undergo impressive outbursts where their luminosity changes by several orders of magnitude, and their spectrum is significantly modified. At least some of this dramatic evolution is associated with changes in the accretion disk \citep{done_modelling_2007}. During these outbursts, the radio emission, a proxy for jet activity, is highly correlated with the X-ray luminosity, a proxy for the accretion rate, linking both processes \citep{corbel_radio/x-ray_2003,corbel_universal_2013}. It is believed that the magnetic field plays a natural role in connecting both mechanisms. Furthermore, the radio luminosity dramatically changes during the outbursts, suggesting that jets turn on and off.} {Models explaining this behavior assume that the vertical magnetic field is advected and diffused during the outburst, mediating the change in jet luminosity \citep{ferreira_unified_2006,marcel_unified_2019}. A mechanism generating a large-scale vertical magnetic field would complement those models, as it could provide a source for the large-scale vertical magnetic field necessary to power the jets.}

Finally, recent polarization observations of Sgr A* during a bright near-infrared flare have emphasized the necessity of a substantial poloidal magnetic field, that has vertical component \citep{jimenez-rosales_dynamically_2020}. This magnetic field component is crucial for matching the periodicity of the flare. Furthermore, observations from the Event Horizon Telescope focused on polarized emissions around the supermassive black hole in M87 find better agreement with a strong and organized poloidal magnetic field \citep{collaboration_first_2021}.


\subsection{The MRI dynamo}
An efficient mechanism that generates a large-scale poloidal magnetic field from the dominant toroidal magnetic field could be the MRI. 
The MRI turbulence was investigated as a dynamo mechanism soon after its discovery \citep{brandenburg_dynamo-generated_1995,hawley_local_1996}. An initially zero net vertical flux magnetic field leads to self-sustaining extended duration turbulence with an effective angular momentum transport coefficient, of at most  $\alpha_\nu\sim0.01$ \citep{shakura_black_1973,fromang_mhd_2007-1}. The turbulence can be sustained because the initial MRI-unstable magnetic field can self-generate and self-sustain \citep{rincon_self-sustaining_2007}. 

This self-sustaining process has been studied in detail in non-stratified shearing box simulations \citep{lesur_self-sustained_2008,herault_periodic_2011,riols_global_2013,riols_dissipative_2015,riols_magnetorotational_2017,mamatsashvili_zero_2020,held_mri_2022}. The self-sustaining process can be broken down into three main ingredients: (1) First, a large-scale axisymmetric toroidal field is generated through Keplerian shear action on a weak large-scale poloidal field. (2) This large-scale toroidal field is unstable to the non-axisymmetric MRI, driving perturbations in the form of MHD shearing waves \citep{goldreich_ii_1965,balbus_powerful_1992,johnson_magnetohydrodynamic_2007}. (3) Non-linear wave interactions generate a large-scale axisymmetric poloidal magnetic field. This large-scale poloidal can then be sheared again into a toroidal field to restart the process. It must be stressed that the MRI dynamo is intrinsically nonlinear and very different from a typical kinematic dynamo; a finite initial seed field is needed to kick-start the whole process \citep{rincon_dynamo_2019}.

In stratified shearing boxes, the MRI dynamo self-organizes into quasi-periodic reversals of the large-scale magnetic field \citep{brandenburg_dynamo-generated_1995,davis_sustained_2010,gressel_mean-field_2010,gressel_characterizing_2015,salvesen_accretion_2016}. These reversals propagate from the disk midplane to the disk corona and look qualitatively similar to the butterfly diagram of the solar dynamo. Recently the self-sustaining process described above was extensively characterized in stratified shearing boxes \citep{held_mri_2023}.
To corroborate the relevance of the MRI dynamo for mergers, the butterfly diagrams was also observed in dynamical general relativistic MHD simulations of BHNS mergers \citep{hayashi_general-relativistic_2022,hayashi_general-relativistic_2023}.


While there has been some discussion of the dynamo-generating mechanism operating in global 3D simulations, most of the work has focused on finding reduced mean-field models \citep{flock_large-scale_2012,hogg_influence_2018,dhang_characterizing_2020}.
While useful, mean field models linearize the dynamics and can not capture the full nonlinear dynamo physics. Furthermore, it is crucial to understand the physical mechanism before extrapolating mean field models to astrophysical regimes.

We hope the analysis we present here can inform the construction of mean-field models that could help computing merger simulations with weaker initial magnetic fields.
The relevance of this mechanism for mergers and disruptions means that a quantitative analysis of the self-sustaining MRI dynamo mechanism in global simulations is long overdue. Recently, \cite{liska_large-scale_2020} simulated an accretion disk with an initially toroidal magnetic field and found that a large-scale poloidal field emerged naturally.  Then, this large-scale poloidal magnetic field accumulates into the central BH until it reaches the magnetically arrested state \citep[MAD, ][]{narayan_magnetically_2003,igumenshchev_three-dimensional_2003,tchekhovskoy_efficient_2011}, in which the vertical magnetic flux is dynamically important. Where MADs can be thought of as the natural end state of magnetized accretion that posseses a large amount of large-scale vertical magnetic flux \citep{jacquemin-ide_magnetic_2021}. 

This manuscript elucidates the self-sustaining MRI dynamo in global 3D general relativistic MHD (GRMHD) simulations of radiatively inefficient accretion flows (RIAFs). This mechanism appears to be a generic feature of thick and highly turbulent MHD accretion disks, making it applicable to a wide range of astrophysical sources. In this study, we specifically concentrate on the mergers of compact objects, as they offer a compelling context for a dynamo mechanism capable of producing large-scale vertical magnetic fields.
The manuscript is organized as follows. Section \ref{sec:allmethods} introduces the numerical experiment and our Reynolds averaging method. Section \ref{sec:quali} qualitatively describes the main features of the numerical experiment and their time evolution. Section \ref{sec:main_dinam} quantitatively describes the magnetic field generation mechanism. Section \ref{sec:clossure} finds the best mean-field model to describe the dynamo mechanism.  Section \ref{sec:conclu_discu}, summarizes and discusses our results.

\section{methods}\label{sec:allmethods}
\subsection{Numerical setup}\label{sec:setup}

We employ the \texttt{H-AMR} code \citep{Liska2022} to solve the set of ideal GRMHD equations on a spherical polar grid ($r$, $\theta$, $\varphi$) in the Kerr-Schild coordinates \citep{gammie_harm_2003}.  
We also define for convenience the cylindrical radius $R$ and the vertical height $z$. We adopt dimensionless units such that $G = M = c = 1$, where $M$ is the mass of the black hole. This implies that in our units, the gravitational radius is unity, $r_{\rm g} = GM/c^2 = 1$. For the magnetic field, we use Lorentz-Heaviside units, such that the magnetic pressure is given by $b^2/2$ in terms of the fluid-frame magnetic field strength, $b$.

In this work, we reanalyze the simulation first presented by \cite{liska_large-scale_2020}.
An accretion disk orbiting a nearly maximally rotating BH ($a=0.9$) is simulated. The disk is initialized with a sub-Keplerian \cite{chakrabarti_1985} torus, where the specific angular momentum profile follows $l\propto R^{1/4}$. The torus has an inner edge at $r_{\rm in} = 6\,r_g$, a pressure maximum at $r_{\rm max} = 13.792\,r_g$, and an outer edge at $r_{\rm out} \approx 4 \times 10^4\,r_g$. The grid extends to $r= 10^5 r_g$ and the inner boundary is causally disconnected from the flow inside the event horizon. We use a polytropic equation of state with $\gamma=5/3$, which gives a torus scale height of $H/r \sim 0.2$ at $r_{\rm max}$ and $\sim0.5$ at $r_{\rm out}$. We use transmissive BCs at the poles, periodic BCs in the $\varphi$-direction, and outflowing BCs at the inner and outer radial boundaries are used.  We use the piecewise parabolic method \citep{Colella} for spatial reconstruction and second-order time-stepping. The simulation uses a base grid of resolution $N_r\times N_{\theta}\times N_{\varphi} = 1872 \times 624 \times 128$ that is uniform in $\log r, \theta$, and $\varphi$, respectively. On top of the base grid, we use 3 levels of static mesh refinement (SMR) in the $\varphi$-direction: this progressively increases the effective $\varphi$-resolution from the pole to equator and maintains the cell aspect ratio close to unity. This leads to an effective resolution, $N_r\times N_{\theta}\times N_{\varphi} = 1872 \times 624 \times 1024$, within 60 degrees of the equator that corresponds to 70-90 cells per disc scale height. The MRI is seeded with a strong large-scale and uniform toroidal field, with an initial plasma beta $\beta_{\rm ini}=2P/b^2=5$; where $P$ is the fluid-frame gas pressure. We also analyze a supplementary simulation identical to the first, except that the toroidal magnetic field polarity is inverted in the south hemisphere (see Appendix \ref{A:sim2}).

An initially axisymmetric toroidal magnetic field might seem like an unrealistic initial condition. However, as described in  Section~\ref{sec:intro} a dominant toroidal magnetic field is naturally generated due to azimuthal shear in dynamical space-time GRMHD simulations after the early collision/disruption phase. Furthermore, \cite{aguilera-miret_role_2023} find that the strong toroidal magnetic fields organize into axisymmetric structures.

\cite{liska_large-scale_2020} evolved the simulation out to $t = 120,000\,\,r_g/c$ to verify that the emergent poloidal field was stable on long lived. The analysis presented here focuses solely on the emergence of the poloidal magnetic field, which happens at $t<10,000\,\,r_g/c$. We record the data snapshots at a cadence of $\Delta t = 10\,r_g/c$. At $r = 20\, r_g$, this provides $\sim 9$ data snapshots per Keplerian time scale, $\Omega_K^{-1}= R^{3/2}$, without the $2\pi$.

\subsection{Averaging procedure}\label{sec:average}
Throughout this manuscript we use a Reynolds decomposition to decompose all quantities as
\begin{equation}
    X = \mean{X}+\de X,
    \label{Eq:Rey_avg}
\end{equation}
where $\mean{X}$ is the average or large-scale component and $\de X$ is the turbulent component. For Eq.~(\ref{Eq:Rey_avg}) to be consistent, the turbulent components must vanish under averaging, $\mean{\de X}=0$. Only nonlinear terms, $\mean{\de X\de Y}$, do not vanish under averaging due to their correlations. Those correlation terms represent the backreaction of the turbulence on the mean fields.

The Reynolds decomposition is defined here with the azimuthal average,
\begin{equation}
    \mean{X} =\frac{1}{2\pi} \int\limits_{0}^{2\pi}X\rm{d}\varphi.
\end{equation}
Consistent with this definition, we will also refer to average quantities as axisymmetrized. The effect of MHD turbulence on large-scale magnetic field generation is perceived by Reynolds averaging the induction equation. 

\begin{figure*}
	\includegraphics[width=0.85\textwidth]{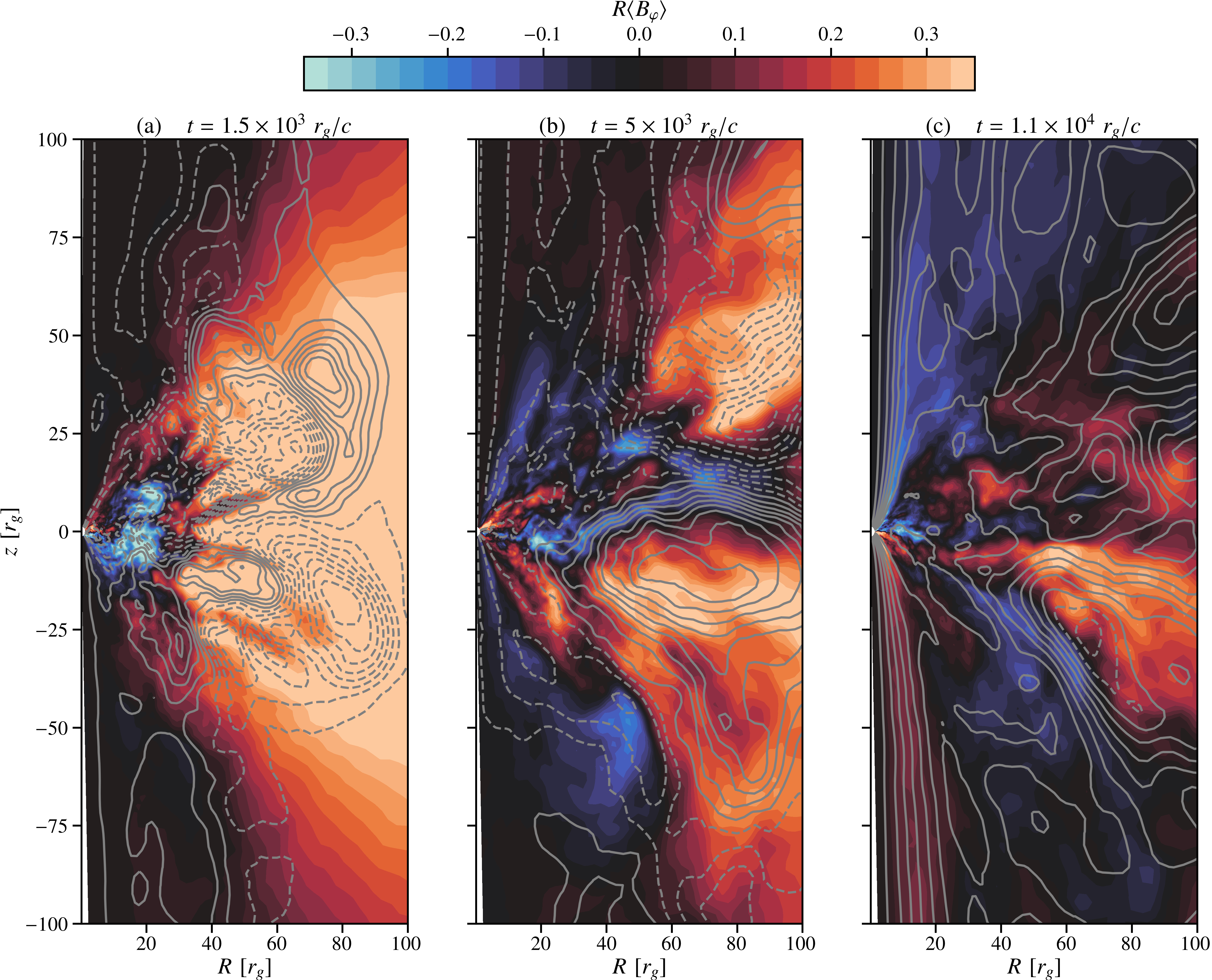}
    \caption{Snapshots of the axisymmetrized toroidal magnetic field, $R\mean{B_\varphi}$, in color as a function of $R$ and $z$. We also show the poloidal magnetic field lines through the poloidal magnetic flux (Eq.~\ref{Eq:pol_flux}): dashed lines show negative polarity and solid lines show positive polarity. The magnetic field structures become larger and larger with time, reflecting the generation of large-scale poloidal magnetic flux.}
    \label{fig:aphi1}
\end{figure*}
We also define a disk average in the latitudinal coordinate, 
\begin{equation}
    X|_{d} =\frac{1}{\int^{\theta_2}_{\theta_1}\sqrt{-g}\rm{d}\theta} \int\limits^{\theta_2}_{\theta_1}\sqrt{-g}\mean{X}\rm{d}\theta,
    \label{Eq:disk_avg}
\end{equation}
where $g$ is the determinant of the metric. The disk average will only be applied to already azimuthaly averaged quantities ($\mean{X}$ or $\mean{XY}$). We choose
\begin{equation}
    \theta_{1,2} = \frac{\pi}{2}\pm \arctan\left(3\frac{h}{R}\right),
\end{equation}
where $h$ is the disk geometrical thickness, which is $h/R\simeq 0.35$ in the regions of interest. This choice of disk average avoids including the jet funnel and includes most of the turbulent signal. We verify that this choice captures most of the turbulent signal in Section \ref{sec:main_dinam}. We experimented with other choices and found identical trends.

We define the poloidal magnetic field as
\begin{equation}
    \mathbf{B_p} = B_r\vek{e}_r+B_\theta\vek{e}_\theta.
\end{equation}
Within this manuscript all quantities will be in physical units unless stated otherwise, thus the polar field above is actually $B_{\hat{\theta}} =\sqrt{g_{\theta\theta}} B^\theta$. Throughout the rest of the manuscript we drop the hats for clarity. 

The Reynolds decomposition can be readily performed on the Newtonian induction equation,
\begin{equation}
    \pdv{\mathbf{B}}{t} = \vgrad{}\times \left(\mathbf{u}\times \mathbf{B}\right),
    \label{Eq:Newt_induc}
\end{equation}
to make the analysis more tractable. As we will see the magnetic field is generated at large distances far from the black hole, where this approximation is accurate. We check that energy is conserved, a posteriori, by comparing the left- and right-hand sides of Eq.~(\ref{Eq:Newt_induc}). If energy is well conserved our Newtonian approximation is accurate at the radii in question and farther away.  The average of Eq.~(\ref{Eq:Newt_induc}) gives,
\begin{equation}
    \pdv{\vmean{B}}{t}=\vgrad{}\times \left(\vmean{u}\times \vmean{B}+\vEMF \right),
\end{equation}
where the gradient now only depends on $r$ and $\theta$. We define the turbulent electromotive force (EMF)
\begin{equation}
    \vEMF = \mean{\de \mathbf{u}\times \de \mathbf{B}}.
    \label{Eq:avg_induct}
\end{equation}
The turbulent EMF is the feedback of turbulence on the mean magnetic field, it dissipates or generates mean magnetic fields and is at the core of large-scale dynamo theory. For magnetic field generation and sustenance to be possible, there needs to be an energy feedback loop connecting the mean toroidal and poloidal magnetic fields.
The turbulent EMF plays a critical role connecting the two magnetic field components.


\section{Description of the simulation}\label{sec:quali}

\begin{figure}
	\includegraphics[width=\columnwidth]{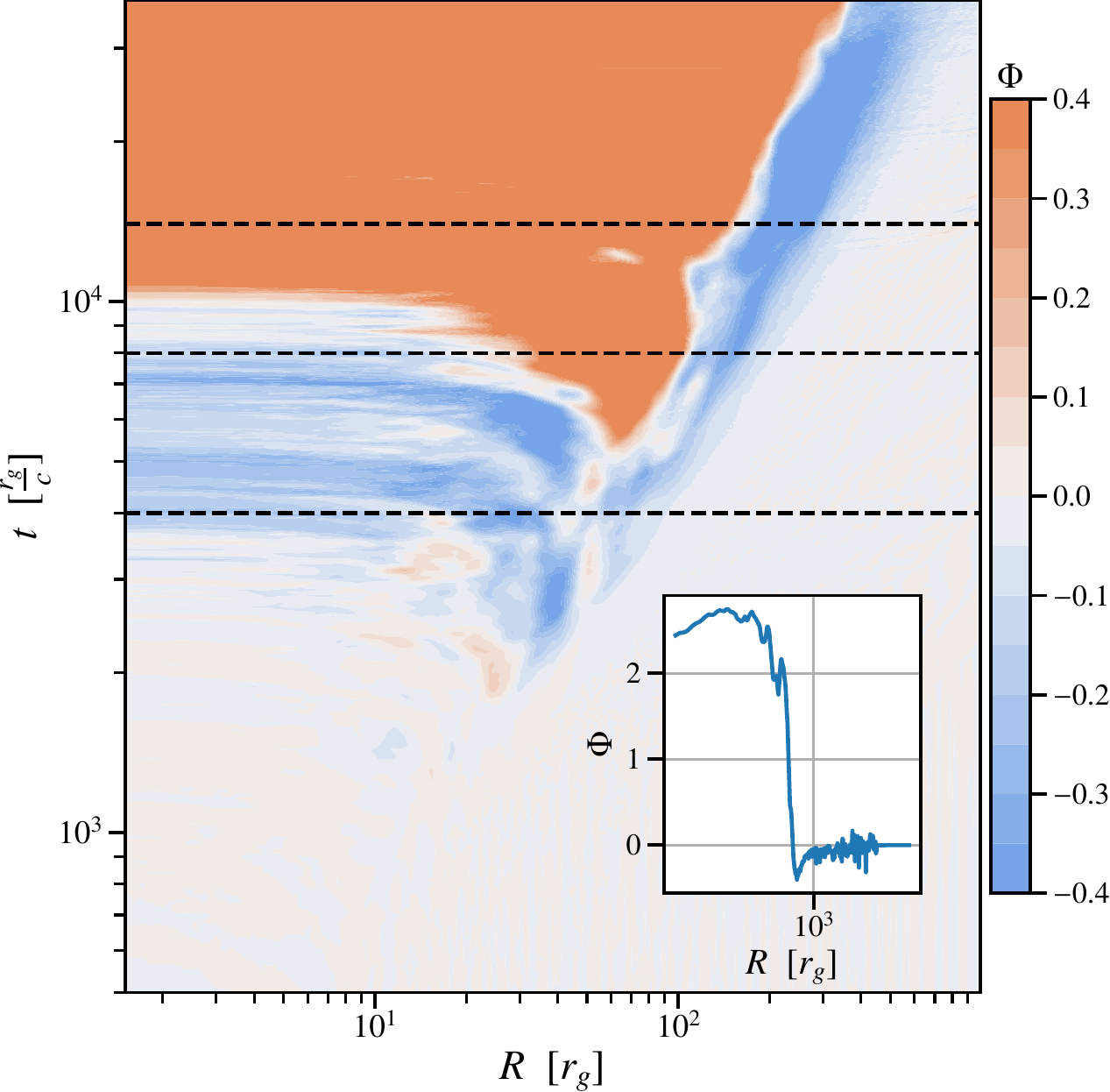}
    \caption{Axisymmetrized poloidal magnetic flux (Eq.~\ref{Eq:pol_flux}), $\Phi$, evaluated at $\theta=\pi/2$ as a function of $t$ and $R$. The vertical dashed lines show the times of the snapshots of Fig.\ref{fig:aphi1}. Notice the large-scale advection of magnetic field structures.
    The dominant large-scale flux loop appears at $R\sim 60r_g$ and $t\sim4.5\times10^{3}r_g/c$.}
    \label{fig:large_loops}
\end{figure}

In this section, we set the stage with a description of the qualitative evolution of the magnetic field during the simulation. We then quantitatively characterize the saturated turbulent state of the simulation.
\subsection{Qualitative evolution of the magnetic field}
Figure \ref{fig:aphi1} shows $R\mean{B_\varphi}$ in color and the poloidal magnetic field lines through the magnetic potential
\begin{equation}
    \Phi(r,\theta,t) = \int\limits\mean{ B^{r}} \sqrt{-g}\diff \theta,
    \label{Eq:pol_flux}
\end{equation}
where $B^r$ is the contravariant component of the radial magnetic field. Figure \ref{fig:large_loops} shows a space-time diagram of the magnetic flux through the disk midplane, $\Phi(r,\theta=\pi/2,t)$. At $t=4000\,\,r_g/c$, the system has already formed axisymmetric poloidal magnetic fields, $\mean{\Bpo}$, in loop structures, see Fig.~\ref{fig:aphi1}(a). The growth of the axisymmetric poloidal field occurs on a relatively short timescale. 

\begin{figure}
	\includegraphics[width=\columnwidth]{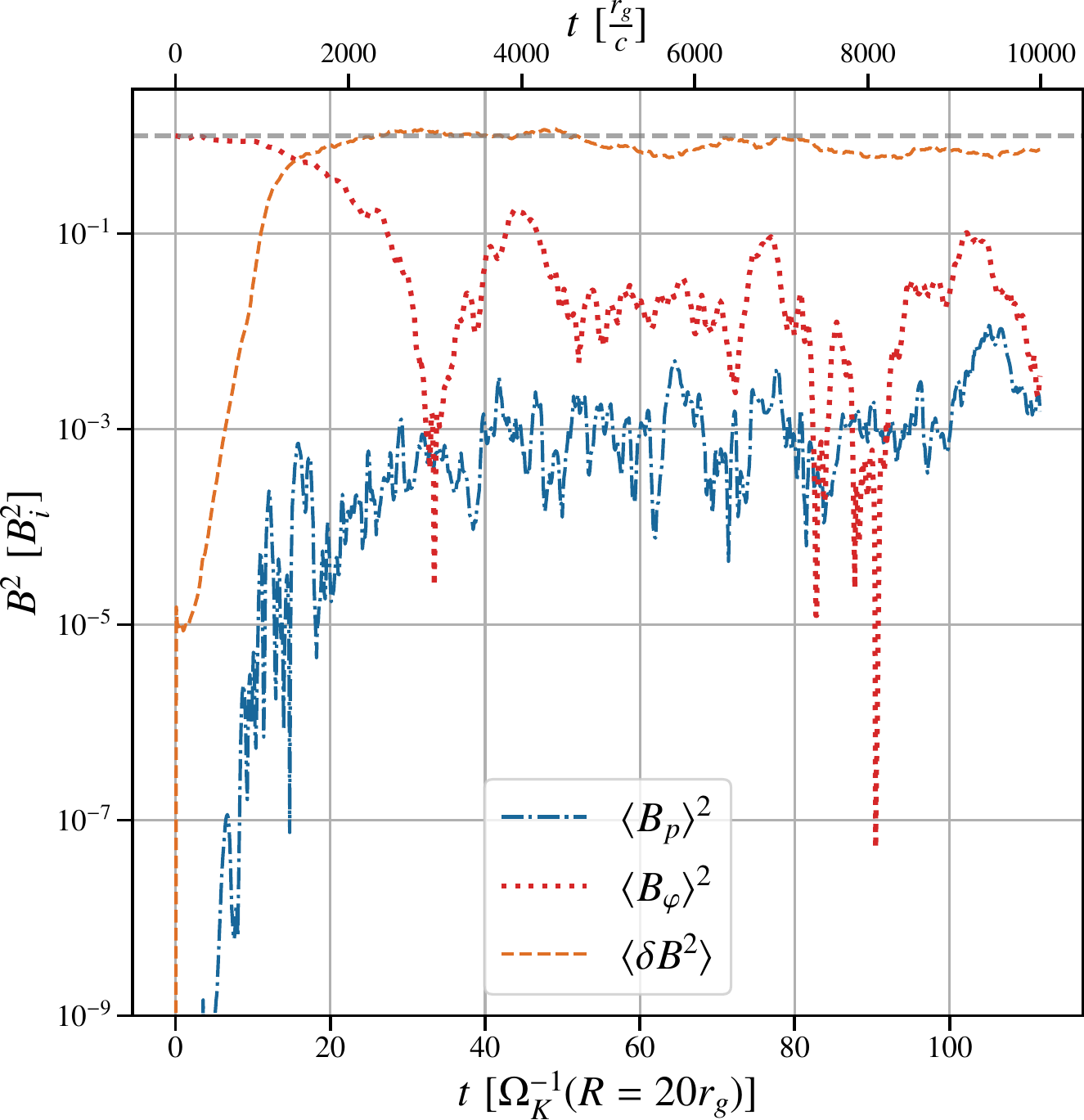}
    \caption{Time-evolution of different axisymmetrized magnetic-field energies averaged in the disk and evaluated at $r=20r_g$. We observe a clear initial exponential growth for $\mean{B_p}^2$ and $\mean{\delta B^2}$. After the fluctuations become of the same order as for the average toroidal field, $\mean{B_\varphi}^2$, the axisymmetric toroidal field energy rapidly decreases.}
    \label{fig:growth}
\end{figure}

In Figure \ref{fig:growth}, we show the local time-evolution of the different magnetic energy components averaged in the disk (see Section \ref{sec:average}), evaluated at $r=20r_g$, as functions of the local shear time. As we will see below, the first large-scale poloidal magnetic loop that connects to the BH is originally advected from $r=20r_g$; subsequent dominant poloidal magnetic field polarities originate from even farther out.
We can distinguish two clear phases: (1) an initial exponential growth of the turbulent field, $\mean{\de B^2}$, and the axisymmetric poloidal field, $\mean{B_p}^2$. This exponential growth lasts up to $t = 10\, \Omega_K(r=20r_g)^{-1}$ when the turbulent magnetic energy, $\mean{\delta B^2}$, reaches saturation. (2) Afterwards, the magnetic fields at the local radius reach a non-linear statistically steady state characterized by no net growth and a well defined average. The axisymmetric poloidal magnetic energy, $\mean{B_p}^2$, saturates simultaneously with the turbulent magnetic component, showing their dynamical link. The axisymmetric toroidal magnetic energy, $\mean{B_\varphi}^2$, decreases once the turbulent magnetic energy becomes comparable to it. The backreaction of turbulence on the seed field then erases the initial condition.

On average, the poloidal loop structures have opposite polarities on opposite sides of the disk midplane (see Fig.~\ref{fig:aphi1}a). However, this trend is only roughly true, and disappears at later times; the system is quite stochastic. Within the inner regions the toroidal magnetic field changes signs through shearing of the poloidal magnetic field loops. 
We notice that at high latitudes above the disk midplane the poloidal field forms elongated structures in the form of escaping flux tubes. These are reminiscent of the magnetic towers described by \cite{lynden-bell_why_2003} and show the vertical shedding of magnetic-field structures.

\begin{figure}
	\includegraphics[width=0.85\columnwidth]{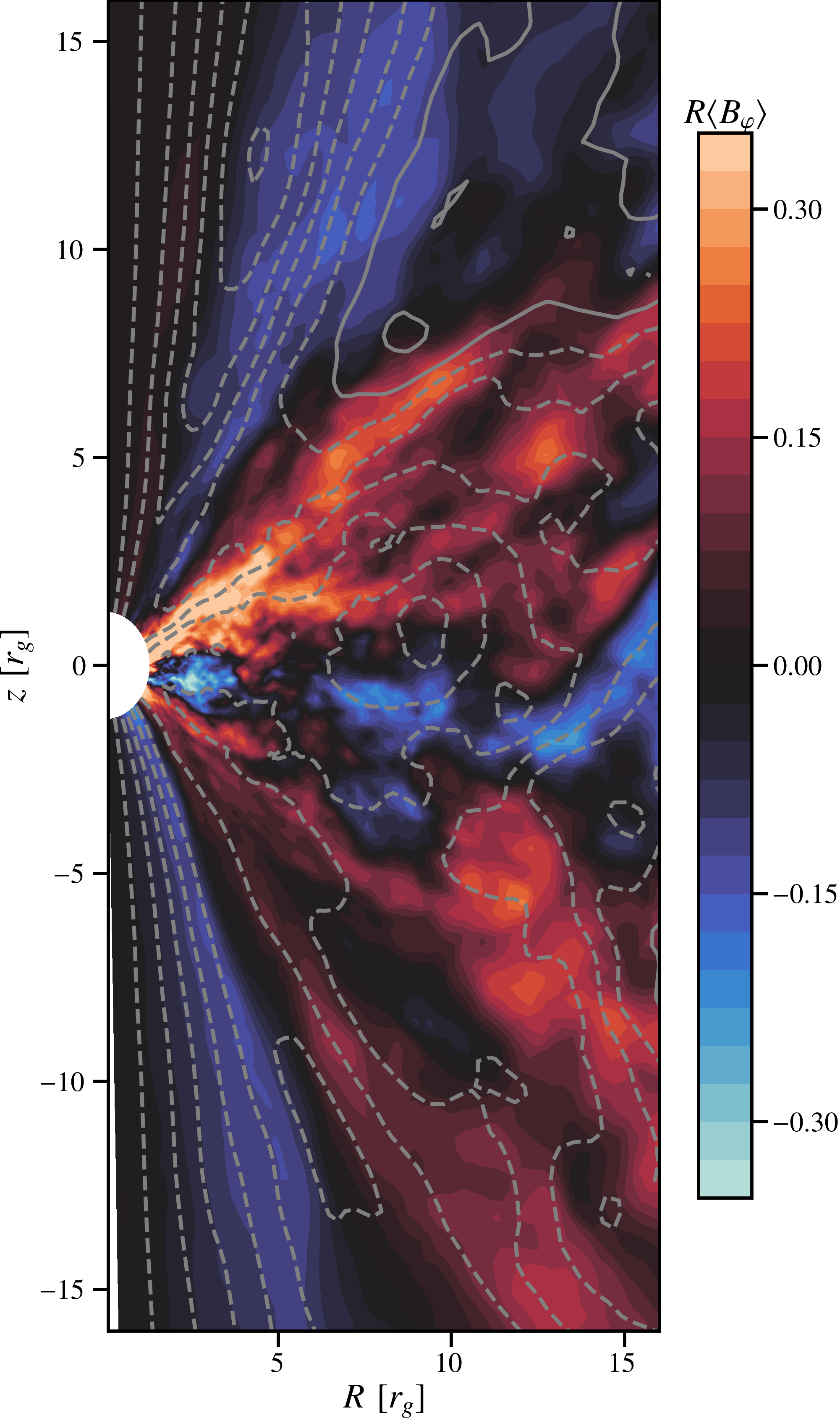}
    \caption{Axisymmetrized toroidal magnetic field, $R\mean{B_\varphi}$ in color as a function of $R$ and $z$ at $t=8\times 10^{3}r_g/c$. We also show the poloidal magnetic field lines through the poloidal magnetic flux (Eq.~\ref{Eq:pol_flux}), dashed lines show negative polarity and solid lines show positive polarity. In the initial stages of evolution, $t<10^{4}r_g/c$, the inner magnetic field structure is distinct from what is typically seen in MADs.}
    \label{fig:zoom_structure}
\end{figure}

For $t<4000\,\,r_g/c$, there is no coherent poloidal field connected to the central black hole but coherent poloidal loops are still present (Fig.~\ref{fig:large_loops}). At $t\sim 4000\,\,r_g/c$ a coherent poloidal structure attaches to the BH. This coherent poloidal loop originates far from the BH, at $r\sim 20r_g$.
We show a zoom in on the axisymmetrized poloidal and toroidal field structures connected to the black hole in Fig.~\ref{fig:zoom_structure}. The topology of how this magnetic field connects to the black hole is distinct from the one observed in magnetically arrested disks \citep[MAD,][]{tchekhovskoy_efficient_2011}. Figure \ref{fig:zoom_structure} shows open and closed field lines, with the closed field lines threading the disk mid plane. This structure drives jets at low energy efficiencies \citep{christie_role_2019,liska_large-scale_2020,gottlieb_large-scale_2023}. A similar structure was studied in 2D GR particle in cell simulations by \cite{el_mellah_spinning_2022}, who showed that it can drive efficient particle acceleration.

The system continues to evolve, for $t>4000\,r_g/c$, and forms loops of larger and larger size; see the movie and Fig.~\ref{fig:aphi1}(b). A large-scale loop emerges at $R\sim 50r_g$ and $t\sim 5000r_g/c$ (see fig.~\ref{fig:large_loops}) and at $t\sim 8000r_g/c$, this poloidal field structure is threading the disk midplane (Fig.~\ref{fig:aphi1}b). The ballooning instability \citep{lynden-bell_why_2003} can then take hold on both hemispheres. This instability expands the poloidal loop, which pushes away all other magnetic loops, leading to a dominant polarity that connects to the black hole, see Fig.~\ref{fig:aphi1}(c).
At $t\sim 10000r_g/c$, the large-scale loop manages to become the dominant polarity close to the BH. The inset in Fig.~\ref{fig:large_loops} representing $\Phi(r,\theta=\pi/2,t)$ at $t\sim3\times10^{4}$ shows that there is not net polarity in the system taken as a whole (i.e., out to large radii). Only in the vicinity of the BH does a net polarity appear. The polarity emergent at $t\sim10^{4}r_g/c$ endures for the rest of the simulation and is still present at $t\sim1.5\times10^{5}r_g/c$ \citep{liska_large-scale_2020}.

Even though the ballooning instability can naturally explain the runaway growth of a magnetic field loop whose vertical length scale, $l_{\rm loop}$, is larger than the disk scale height, $h$, it cannot explain how a poloidal magnetic field larger than the disk scale height appears in the first place. Furthermore, it is surprising that the loop that connects to the BH at $t\sim 4000r_g/c$ does not become the dominant polarity, while the one that connects at $t\sim 10^{4}r_g/c$ does. In Section~\ref{sec:main_dinam} we discuss the mechanism that generates the poloidal magnetic field and how the dominant polarity takes hold.



\begin{figure}
	\includegraphics[width=\columnwidth]{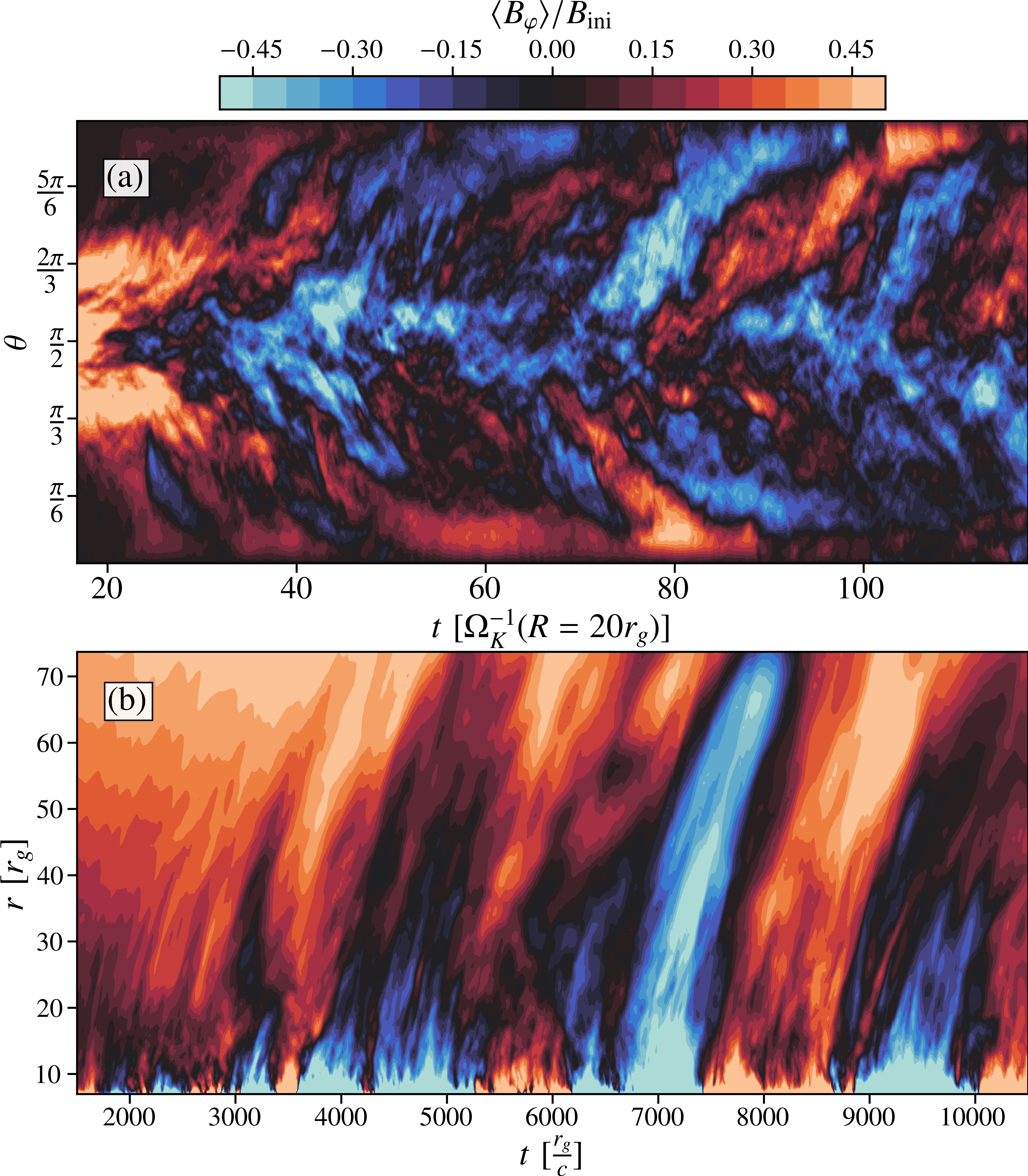}
    \caption{{\bf [Panel (a)]:} $\theta-t$ diagram of the axisymmetrized toroidal magnetic field, $\mean{B_\varphi}$, evaluated at $r =20r_g$, normalized to the initial toroidal field. We observed the ejection of inverting toroidal structures, reminiscent of the typical butterfly diagram of MRI turbulence. However, the structures in our simulation are more stochastic than those typically seen in MRI turbulence. {\bf [Panel (b)]:} $r-t$ diagram for the axisymmetrized toroidal magnetic field, $\mean{B_\varphi}$, evaluated at the surface of the disk, $\theta = \pi/2+\arctan(3h/R)$, normalized to the initial field. Large-scale (as large as $\sim 70r_g$) toroidal field structures of differing polarity are ejected from the system. There are inversions in the polarity of the field, but no clear periodicity is observed. }
    \label{fig:transport_tor}
\end{figure}

\subsection{Butterfly diagram}

Figure~\ref{fig:transport_tor}(b) shows a $\theta$-$t$ diagram, also called the butterfly diagram, of the toroidal magnetic field at $r=20r_g$. The toroidal magnetic field is ejected from the system and the ejection happens in cycles with polarity inversions. Shearing box simulations show similar flux inversions with a vertical propagation \citep{gressel_characterizing_2015}. However, periodic cycles in shearing box simulations are far more regular than those observed in Fig.~\ref{fig:transport_tor}(a).  
This is consistent with \cite{hogg_influence_2018}, who showed that thicker disks have a more incoherent dynamo cycle.

In Fig.~\ref{fig:transport_tor}(a) we show an $r$-$t$ diagram for the toroidal magnetic field evaluated at the surface of the disk, $\theta = \pi/2+\arctan(3h/R)$. We see magnetic flux inversions, as in the butterfly diagram. However, in this case the inversions show a certain degree of coherence for multiple radii. We believe that these eruptions of magnetic field are due to the ballooning instability discussed above. As the poloidal magnetic field loops grow, they can shed larger loop structures vertically. This leads to the radially coherent field structures seen in Figure~\ref{fig:transport_tor}(b).  

\subsection{Turbulent steady-state of the system}

\begin{figure}
	\includegraphics[width=\columnwidth]{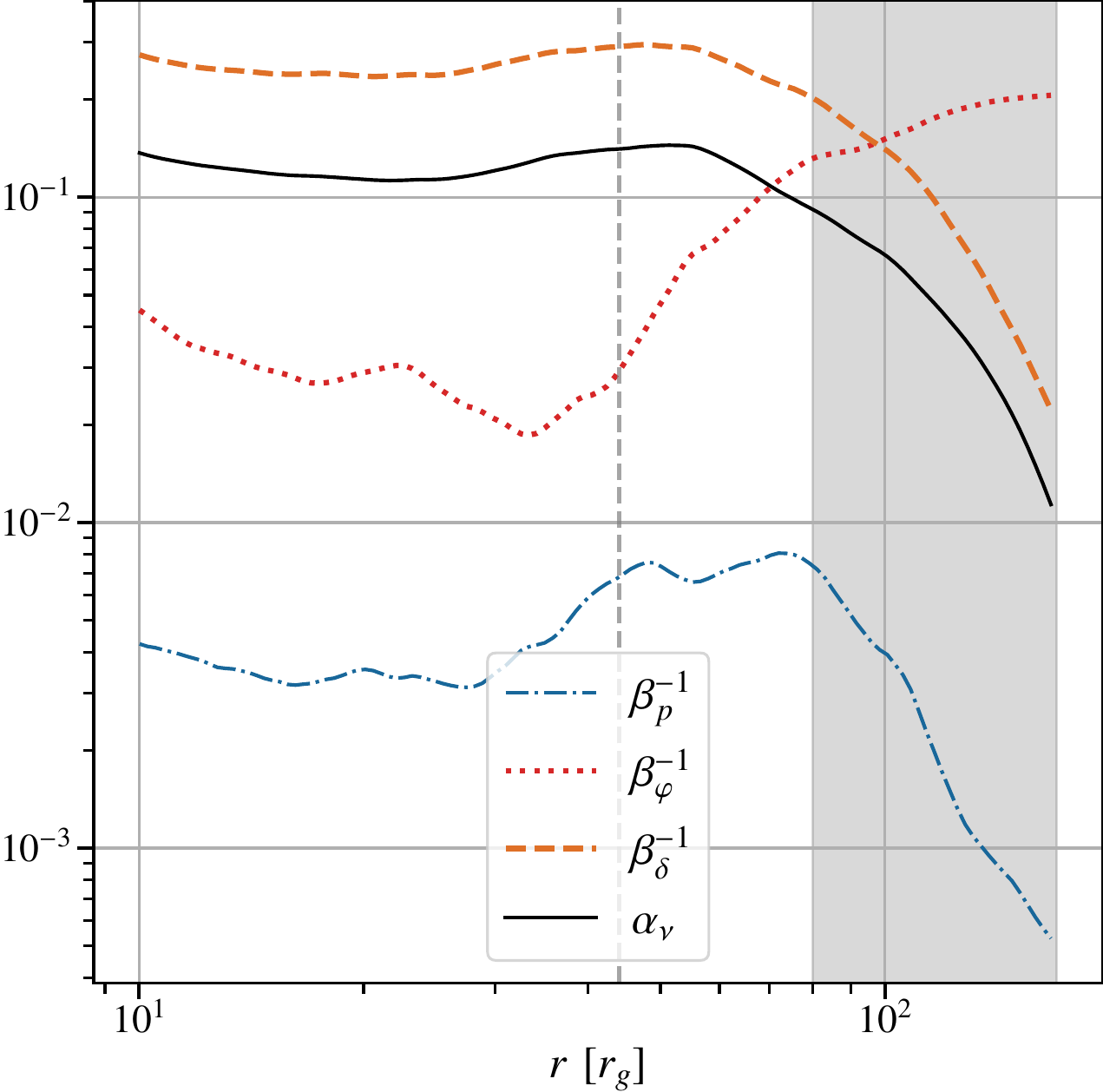}
    \caption{Inverse plasma betas for the different magnetic field components, $ \beta_\delta^{-1}$ for the turbulent field, $\beta_p^{-1}$ for the axisymmetrized poloidal field, $\beta_\varphi^{-1}$ for the axisymmetrized toroidal field, and the angular momentum transport $\alpha_{\nu}$ coefficient. They are vertically averaged in the disk and temporally averaged between $t=5000\,\, r_g/c$ and $t=8000\,\,r_g/c$. The shaded regions highlights the non-saturated outer radii, $r>70r_g$. For $r<70r_g$ the plasma betas are relatively constant and have reached a saturated quasi steady state. The vertical dashed line highlights the point where $\mean{u_r}=0$, as shown in Fig.~\ref{fig:Radius_energyPo}b.}
    \label{fig:Rad_beta}
\end{figure}

In the quasi-statistically-steady state, we have an energetic ordering, $\mean{\delta B^2}\gg \mean{B_\varphi}^2\gg \mean{B_p}^2$, with about an order of magnitude difference between each component (see Fig.~\ref{fig:growth}). The turbulent magnetic energy saturates at roughly the energy of the initial condition. We define different plasma betas for every energetic component
\begin{align}
    \beta_\delta &= \frac{8 \pi \mean{P}}{\mean{\delta B^2}},\\
    \beta_p &= \frac{8 \pi \mean{P}}{\mean{ B_p}^2},\\
    \beta_\varphi &= \frac{8 \pi \mean{P}}{\mean{ B_\varphi}^2},
\end{align}
and the average turbulent stress
\begin{equation}
    \alpha_{\nu} = -\frac{\mean{\delta B_\varphi \delta B_r}}{4 \pi \mean{P}},
\end{equation}
where we only compute the Maxwell stress because the Reynolds stress is negligible in our simulations. In Fig.~\ref{fig:Rad_beta}, we show the inverse of the different plasma betas and the turbulent stress vertically averaged in the disk (Eq.~\ref{Eq:disk_avg}) and temporally averaged between $t=5000\,\, r_g/c$ and $t=8000\,\,r_g/c$. The vertical average is computed after squaring the magnetic field or else opposite polarity fields would cancel out, leading to an artificially small measurement (see Fig.\ref{fig:aphi1}). The shaded region highlights the regions that have not converged and do not follow the same trend as the inner radii.

We measure $\beta_p\simeq 300$ and $\beta_\varphi \simeq 40$.  The toroidal field lost an order of magnitude in magnetic energy; most of that energy was converted into or dissipated by the turbulent fluctuations, which saturate at a $ \beta_\delta \simeq 4$, of the order of $\beta_{ \rm ini}=5$. At large radii, $r>50 r_g$, the different components trend towards their initial values ($\beta_\delta^{-1} \xrightarrow{} 0$, $\beta_p^{-1} \xrightarrow{} 0$, $\beta_\varphi^{-1} \xrightarrow{} \beta_{\rm ini}^{-1}$) as the dynamo has not reached saturation at those radii. 

We measure an $\alpha_{\nu}\simeq 0.1$ a factor of $10$ larger than in typical zero-net flux shearing box simulations \citep{hawley_local_1996}. We attribute this large value to the initially strong toroidal field. \cite{hawley_local_1995} measured the following scalings for fully developed MRI turbulence in shearing boxes with an initially toroidal field \citep[see also 8.1.1. in][]{lesur_magnetohydrodynamics_2020}
\begin{align}
    \label{eq:alpha_bet}
    \alpha_{\rm H95} &\simeq 0.51 \beta_\delta^{-1},\\
    \label{eq:alpha_mean}
    \alpha_{\rm H95} &\simeq 0.35 \beta_{\rm ini}^{-1/2}.
\end{align}
For our initial condition Eq.~(\ref{eq:alpha_mean}) gives $\alpha_{\rm H95}\simeq 0.16$ which is remarkably close to the value we compute. Equation (\ref{eq:alpha_bet}) implies a factor of $2$ difference between $\beta_\delta ^{-1}$ and $\alpha_{\nu}$ which can be read from Fig.~\ref{fig:Rad_beta}.
Overall, we find that the saturated turbulent configuration of our simulation is consistent with scalings derived from shearing box simulations of MRI.
\section{Analysis of the dynamo mechanism}\label{sec:main_dinam}
We now proceed to analyze the dynamical mechanism underlying the magnetic evolution described above.

\subsection{Toroidal magnetic field budget}
To better track the dynamics and sources of magnetic energy we perform a Reynolds average decomposition of the magnetic energy induction equation. First, we dot Eq.~(\ref{Eq:avg_induct}) with the average toroidal field to find,
\begin{equation}
    \frac{1}{2}\pdv{\mean{B_\varphi}^2}{t} = \mathcal{S}_\varphi+\mathcal{A}_\varphi+\delta\mathcal{A}_\varphi,
    \label{Eq:tor_ener}
\end{equation}
where
\begin{equation}
    \mathcal{S}_\varphi = \vmean{B_\varphi}\cdot\rot\left(\vmean{u_\varphi}\times\vmean{B_p}\right),
\end{equation}
is the stretching of poloidal into toroidal field by the large-scale shear ($\Omega$-effect),
\begin{equation}
    \mathcal{A}_\varphi = \vmean{B_\varphi}\cdot\rot\left(\vmean{u_p}\times\vmean{B_\varphi}\right),
\end{equation}
describes the large-scale axisymmetric advection of toroidal field, and
\begin{equation}
    \de \mathcal{A}_\varphi = \vmean{B_\varphi}\cdot\rot\left(\mathbf{\mathcal{E}_p}\right),
\end{equation}
captures the turbulent generation or dissipation, depending on the sign, of axisymmetric magnetic field.
\begin{figure*}
	\includegraphics[width=0.7\textwidth]{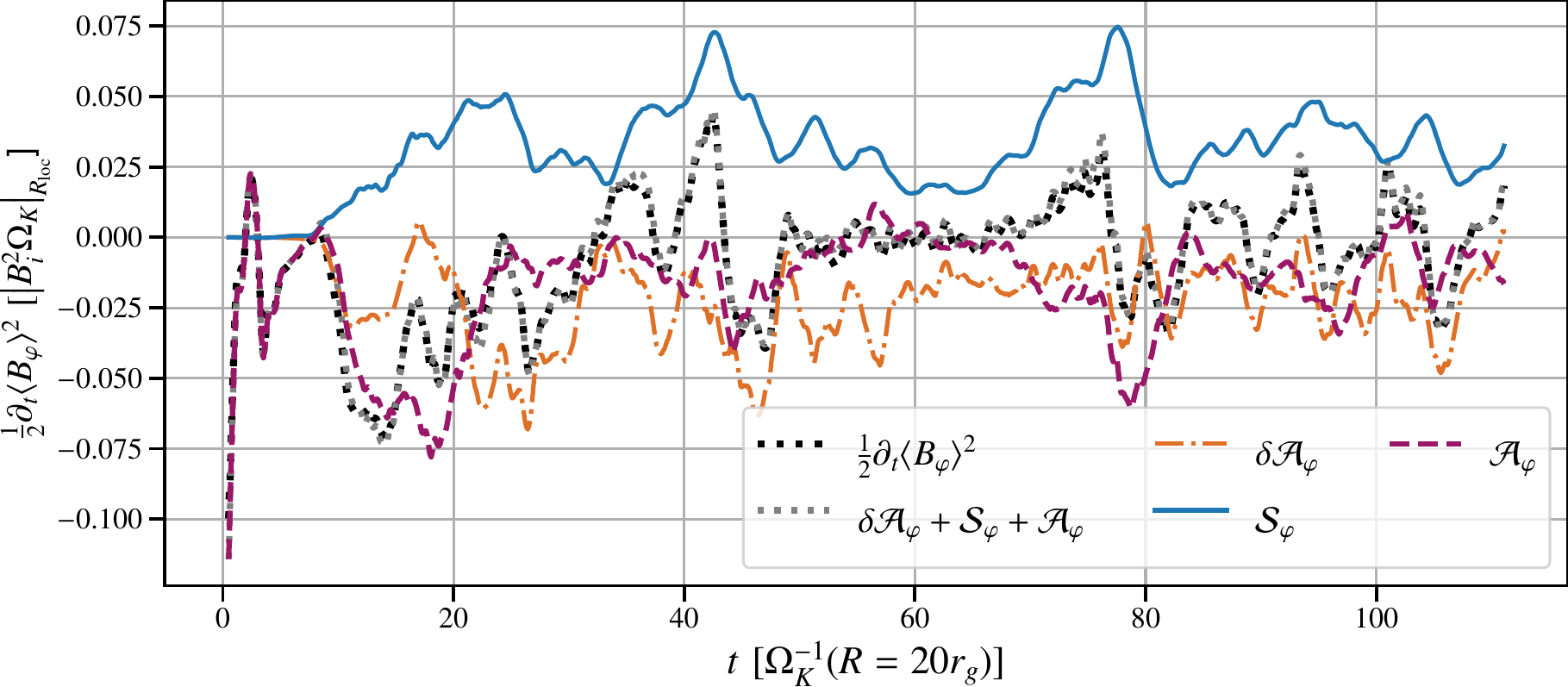}
    \caption{ Different contributions to the toroidal magnetic energy equation (Eq.~\ref{Eq:tor_ener}), averaged within the disk and evaluated at $r=20r_g$, and are normalized to the initial magnetic energy and the Keplerian disk frequency at that radius.  Toroidal magnetic energy is only generated through shear of the poloidal field, $\mathcal{S}_\varphi$. The turbulent component, $\delta \mathcal{A}_\varphi$, always dissipates the toroidal field through turbulent mixing while the contribution of the, poloidal fluid motions, $\mathcal{A}_\varphi$, is to eject the toroidal field from the disk. That the black and gray dotted lines agree, validates the accuracy of our Newtonian approximation and Reynolds decomposition (eq.~\ref{Eq:tor_ener}).}
    \label{fig:To_energy}
\end{figure*}

We show the time dependence of the different, vertically-averaged, components of Eq.~(\ref{Eq:tor_ener}) evaluated at $r=20\,r_g$ in Fig.~\ref{fig:To_energy} (the different terms in Fig.~\ref{fig:To_energy} are vertically-averaged within the disk and normalized to the initial toroidal magnetic field and the Keplerian frequency). The right and left-hand sides of Eq.~(\ref{Eq:tor_ener}) are observed to be equal, validating the use of the Newtonian approximation and showing that our Reynolds decomposition is accurate.

We notice that after $10\Omega_k^{-1}(r=20r_g)$ the system reaches a quasi-statistically-steady state, consistent with Fig.~\ref{fig:growth}.
The only component that produces toroidal magnetic energy is the shear, as in shearing box simulations of the MRI dynamo \citep{lesur_self-sustained_2008,riols_magnetorotational_2017}. The turbulent term, $\de\mathcal{A}_\varphi$, always dissipates toroidal magnetic energy. Shearing box simulations also find that turbulence always dissipates toroidal magnetic energy via turbulent mixing. The advective term, absent in shearing boxes, transports toroidal magnetic energy outwards.

\begin{figure}
	\includegraphics[width=\columnwidth]{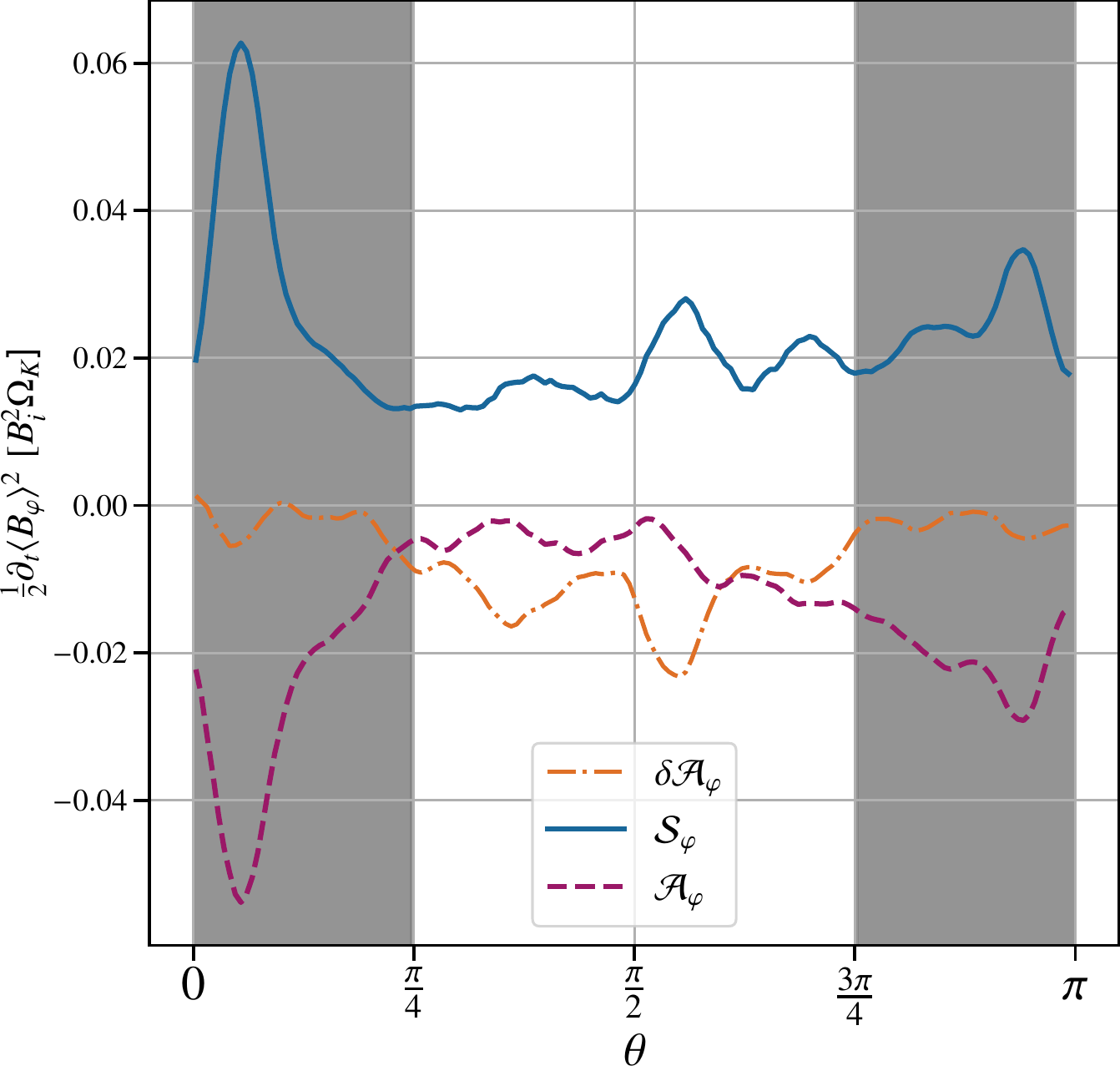}
    \caption{Different contribution to the toroidal magnetic energy equation (Eq.~\ref{Eq:tor_ener}) averaged between $20\Omega_k^{-1}(r=20r_g)$ and $110\Omega_k^{-1}(r=20r_g)$ and evaluated at $r=20rg$, and normalized to the initial magnetic energy and the Keplerian disk frequency at that radius. The shaded regions show $|\theta-\pi/2|>\arctan\left(3\frac{h}{R}\right)$. The toroidal field loses energy to the advection term, $\mathcal{A}_\varphi$, vertically escaping toroidal fields, and the turbulent dissipation, acting within the bulk of the disk. Both terms are cumulatively balanced by the $\Omega$-effect, $\mathcal{S}_\varphi$. }
    \label{fig:eneg_avgTo}
\end{figure}

To better understand the vertical profile of the magnetic field source terms, we compute the time average of the different components of Eq.~(\ref{Eq:tor_ener}) and evaluate them at $r=20$. We average them between $20\Omega_k^{-1}(r=20r_g)$ and $110\Omega_k^{-1}(r=20r_g)$ and show them as functions of $\theta$ in Fig.~\ref{fig:eneg_avgTo}, normalized to the initial magnetic energy and Keplerian frequency. The turbulent magnetic component is only important on the disk mid-plane and tends to 0 outwards. The shaded regions in Fig.~\ref{fig:eneg_avgTo} show the bounds of the disk average defined in Section \ref{sec:average}. We see that our choice of integral bounds includes most of the turbulent signal.

Overall, the $\Omega$-effect compensates for the dissipation due to turbulence and outward advection present within the body of the disk. Above the disk, in the shaded region, shear and avection balance each other. The escaping loops of toroidal magnetic field are clearly visible in Fig.~\ref{fig:transport_tor}. 
 
\subsection{ Poloidal magnetic field budget}
We have found that the sustainment of toroidal field is only possible thanks to the shearing of a large-scale poloidal field. Thus, the dynamics and generation of the poloidal field is essential to the dynamo mechanism as a whole.
In particular, since the only term that produces toroidal magnetic energy is the $\Omega$-effect, poloidal field generation is necessary for toroidal field production above the disk.

\begin{figure*}
	\includegraphics[width=0.7\textwidth]{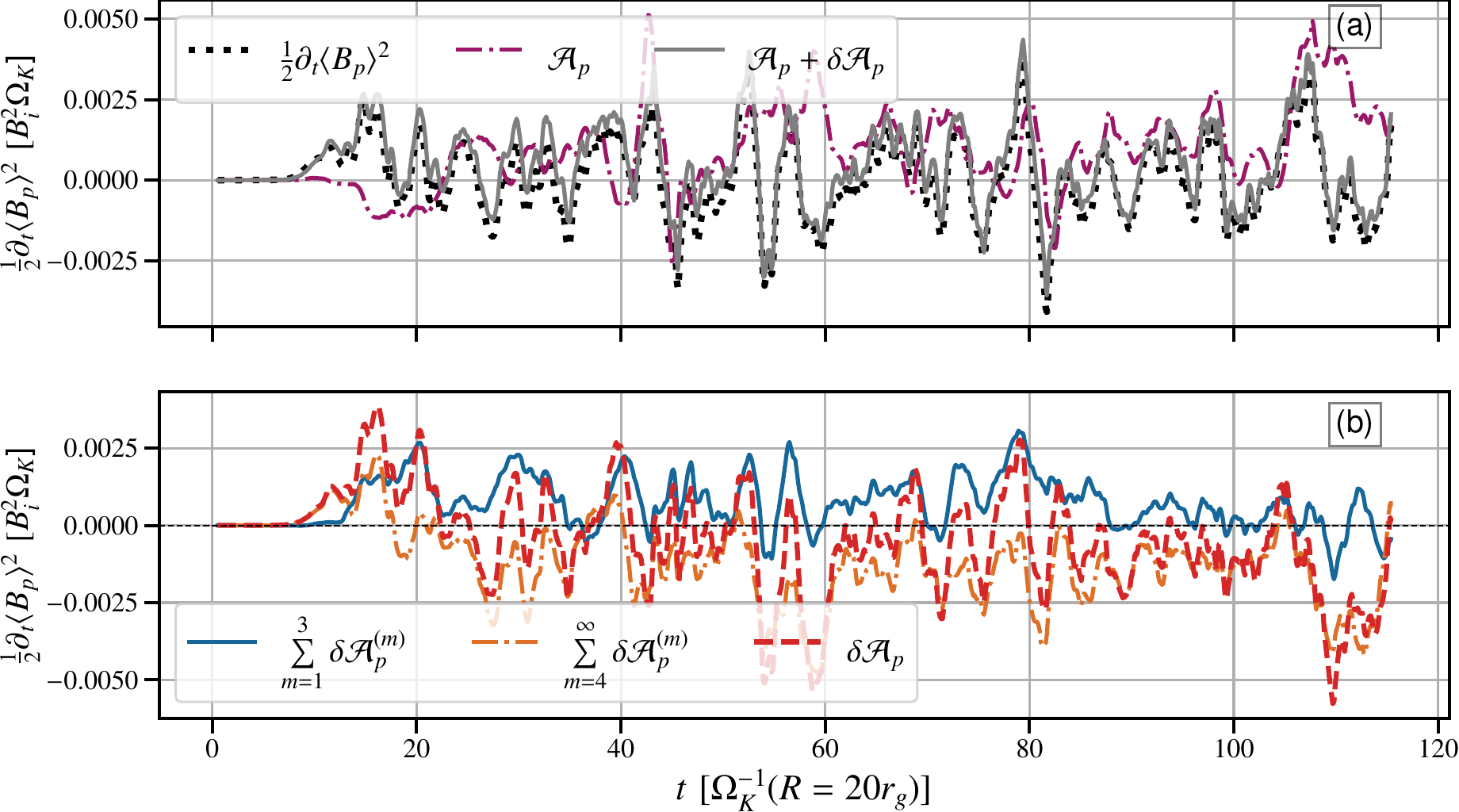}
    \caption{{\bf [Panel (a)]:} Different contributions to the poloidal magnetic energy equation (Eq.~\ref{Eq:pol_energ}) averaged within the disk and evaluated at $r=20r_g$, and normalized to the initial magnetic energy and the Keplerian disk frequency at that radius. {\bf [Panel (b)]:} The sum, $\delta \mathcal{A}_p$, (in red) of magnetic field-producing (or active, shown in blue), $\sum\limits_{m=1}^{3}\de\mathcal{A}_p^{(m)}$,  and dissipative (or passive, shown in orange), $ \sum\limits_{m=4}^\infty\de\mathcal{A}_p^{(m)}$,  modes. The sum of active modes is always positive, while the passive term is mostly negative. Large-scale non-axisymmetric structures, $m=1,2,3$, therefore generate axisymmetric poloidal magnetic fields through their non-linear self interactions. }
    \label{fig:energy_po}
\end{figure*}

We derive an equation for the energetics of the axisymmetrized poloidal field 
\begin{equation}
    \frac{1}{2}\pdv{\mean{B_p}^2}{t} = \mathcal{A}_p+\delta\mathcal{A}_p,
    \label{Eq:pol_energ}
\end{equation}
where
\begin{equation}
    \mathcal{A}_p = \vmean{B_p}\cdot\rot\left(\vmean{u_p}\times\vmean{B_p}\right)
\end{equation}
describes the large-scale or axisymmetric advection of poloidal field, and
\begin{equation}
    \de \mathcal{A}_p = \vmean{B_p}\cdot\rot\left(\mathbf{\mathcal{E}_\varphi}\mathbf{e_\varphi}\right)
\end{equation}
represents the turbulent generation or dissipation of poloidal magnetic field.

In Fig.~\ref{fig:energy_po}(a), we show the different components of Eq.~(\ref{Eq:pol_energ}). We average all components within the disk and evaluate them at $r=20r_g$. We normalize them to the initial magnetic energy and the Keplerian disk frequency at that radius. We notice that our Newtonian Reynolds decomposition is also accurate for the poloidal magnetic energy equation. At $r=20r_g$ the large-scale advection of the magnetic field is positive, i.e. it locally brings poloidal magnetic energy. We explore below how this quantity varies with radius. We show the turbulent EMF term in Fig.~\ref{fig:energy_po}(b). This turbulent term is incoherent; it violently changes signs, producing and dissipating large-scale magnetic energy.
To understand the mechanism generating the magnetic field, we must go a step further, and make a Fourier decomposition of the different modes contributing to the turbulent magnetic production/dissipation.
Using Parseval's theorem we decompose the turbulent term as follows
\begin{equation}
    \delta \mathcal{A}_p = \sum\limits_{m=1}^\infty\de\mathcal{A}_p^{(m)},
    \label{eq:parseval}
\end{equation}
where
\begin{equation}
    \de\mathcal{A}_p^{(m)} = \vmean{B_p}\cdot\rot\left(\mathcal{E}_\varphi^{(m)}\mathbf{e_\varphi}\right),
    \label{eq:Ap_mode}
\end{equation}
and 
\begin{equation}
   \mathcal{E}_\varphi^{(m)}\mathbf{e_\varphi} = \mathcal{R}\left[\mathbf{u_p}^{(m)}\times \mathbf{B_p}^{(m)}|^{*}\right],
    \label{eq:EMF_mode}
\end{equation}
$\mathcal{R}$ represents taking the real part and $X|^{*}$ is the complex conjugate of $X$. Note that $\mathcal{E}_\varphi^{(m)}$ represent the axisymmetric effect of nonlinear interaction of non-axisymmetric $m$-modes.

This procedure is inspired by work done in shearing box simulation, where only the large-scale, low $m$ modes contribute to poloidal field production \citep{lesur_self-sustained_2008,riols_dissipative_2015}.
Here too we measure that only the largest-scale modes, $m=1,2,3$, generate magnetic energy at all radii; we refer to these modes as active (see also Appendix \ref{A:multiwaves}). 
All modes with $m>3$ dissipate magnetic energy on average, we refer to them as turbulent dissipative or passive. In Fig.~\ref{fig:energy_po}b, we show the sum of active, magnetic field-generating, $\sum\limits_{m=1}^3\de\mathcal{A}_p^{(m)}$ vs. dissipative/passive modes, $\sum\limits_{m=4}^\infty\de\mathcal{A}_p^{(m)}$.

The sum of active modes always generates axisymmetric poloidal magnetic fields. The passive modes have a more complicated behavior, generating magnetic energy for $t<20\Omega_k^{-1}$ but dissipating it for $t>20\Omega_k^{-1}$. The time scale when this term goes from generating to dissipating is equal to the saturation time scale shown in Fig.~\ref{fig:growth}. Therefore, we infer that this early behavior is related to the initial linear growth phase of such modes, while the later one is associated with their dissipating-action through a nonlinear turbulent cascade. Due to the short duration of this phase, it does not affect the general trends; the $m>3$ modes still dissipate magnetic energy on average. Furthermore, for $t<20\Omega_k^{-1}$ the active component ($m<3$) is comparable or larger than the turbulent passive component. Therefore, we conclude that the active modes generate basically all the poloidal magnetic energy.

We have also computed the mode decomposition for the toroidal field, shown in Appendix \ref{A:multiwaves}. All modes with $m\geq1$ dissipate toroidal magnetic energy. Thus, the toroidal field loses energy to the active modes, $m=1,2,3$. We also show in Appendix \ref{A:multiwaves} that the active modes transport angular momentum. We can therefore interpret the active modes as MRI-driven perturbations mediated by the axisymmetrized toroidal field.

\begin{figure*}
	\includegraphics[width=0.85\textwidth]{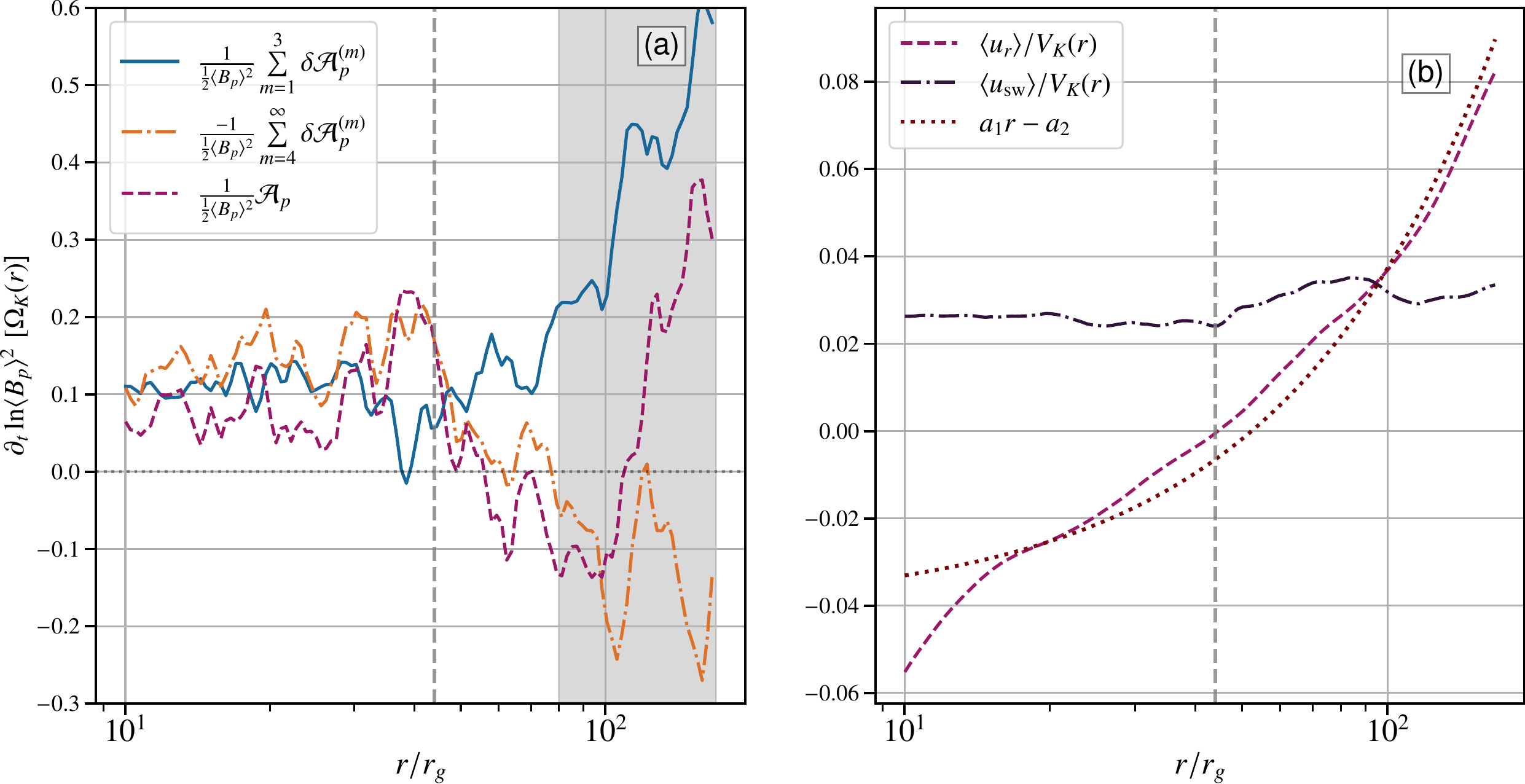}
    \caption{{\bf [Panel (a)]:} Vertically and temporally averaged poloidal energy equation (Eq. \ref{Eq:pol_energ}), normalized to the local Keplerian frequency. The temporal average is performed between $5000r_g/c$ and $8000r_g/c$. The different terms of Eq.~( \ref{Eq:pol_energ}) are divided by the vertical and temporal average of $\mean{B_p}^2$. The opposite of the magnetic energy dissipation term is shown. Advection, $\mathcal{A}_p$, and dissipation, $ \sum\limits_{m=4}^\infty\de\mathcal{A}_p^{(m)}$, of axisymmetric poloidal field balance each other in the inner regions. The vertical dashed line highlights the point where $\mean{u_r}=0$. {\bf [Panel (b)]}: Vertically and temporally averaged radial velocity $\mean{u_r}/V_K$, notice that the trends of $\mathcal{A}_p$ are well explained by the behavior of the radial velocity (see text). We also show an afine fit of the normalized radial velocity. Finally, we show the magnitude of the magnetic field generating velocity, $u_{\rm sw}$. }
    \label{fig:Radius_energyPo}
\end{figure*}

We now look at the large-scale global behavior of the poloidal energy equation (Eq. \ref{Eq:pol_energ}) by vertically and temporally averaging it. The temporal average is performed between $5000r_g/c$ and $8000r_g/c$.
Figure \ref{fig:Radius_energyPo}a shows the vertical average of the different terms of Eq.~(\ref{Eq:pol_energ}) divided by the vertical average of $\mean{B_p}^2$ and normalized to the local Keplerian frequency. Note that here, we show the opposite of the magnetic energy dissipation term (e.g.  multiplied by $-1$). The three terms have similar magnitudes, $~0.1\Omega_K^{-1}$, and are constant for $r<50$ but follow different trends at large radii. First, let us note that in the inner regions of the disk, $r<50$, their signs are coherent with Fig.~\ref{fig:energy_po}. 
The magnetic generation term is roughly constant for the inner radii, only increasing for large radii, $r>70$. We attribute this increase to a lack of saturation in the outer regions, leading to a lower average poloidal magnetic energy (see Fig.~\ref{fig:Rad_beta}). By dividing by the poloidal magnetic energy, we can interpret the magnitude of the quantity, $2{\sum\limits_{m=1}^{3}\mathcal{A}_p^{m}}/{\mean{B_p}^2}$, as the typical frequency of magnetic field regeneration. We define a regeneration time scale,
\begin{equation}
    t_{\rm g} = \frac{\frac{1}{2}\mean{B_p}^2}{\sum\limits_{m=1}^{3}\mathcal{A}_p^{m}}\simeq 10\Omega_K^{-1}.
    \label{Eq:growth_rate}
\end{equation}
This timescale is consistent with Fig.~\ref{fig:growth} up to a factor of $\sim2$, it is also roughly consistent with the cycles of Fig.~\ref{fig:transport_tor}.
The timescale, $t_{\rm g}$, should not be interpreted as a growth rate, as it possesses more information than a growth rate, and encapsulates the saturation energy of the poloidal field.

The large-scale axisymmetric poloidal advection term is positive in the inner regions of the disk, consistent with the inward advection of the poloidal magnetic field (Fig.~\ref{fig:Radius_energyPo}a). We have checked that the advection term, $\mathcal{A}_p$, is dominated by radial advection and  radial velocity divergence.
In Fig.~\ref{fig:Radius_energyPo}(b), we show the disk and temporal average (between $5000r_g/c$ and $8000r_g/c$) of the radial velocity, $\mean{u_r}$. This velocity is negative in the inner regions, consistent with inward advection.

 At the distance, $r_{\rm rev}\sim 50$, the radial velocity changes sign. The inversion of the radial velocity is a feature of our torus initial condition. In standard accretion theory, the radial velocity changes sign wherever $\partial_R \left(R^2\mean{P}\alpha_{\nu}\right)$  changes sign. Hence, this feature is related to our initial pressure maximum. The radii at which we have $\mean{u_r}=0$, viscously move out over the course of our simulation, starting at $r_{\rm rev}=13r_g$ and moving out to $r_{\rm rev}\sim 50$ by $t\sim6\times10^{3}r_g/c$.

At $r \simeq r_{\rm rev}\sim 50$, the advection term, $\mathcal{A}_p$, begins to trend towards negative values. As expected, the advective flux depends on the radial velocity. The advection term is negative for radii $50 < r < 100$, even though the radial velocity is positive at those radii. This is because the radial dependency of the radial velocity avoids a pile-up of the magnetic field in those regions. Magnetic field bundles farther away from $r_{\rm rev}$ travel faster than bundles closer to $r_{\rm rev}$, leading to a chase. The chase of magnetic bundles leads to a net depletion of the magnetic field at those radii, $50 < r < 100$.

In the inner regions, $r<50$, the opposite of the turbulent dissipation term ($-\sum_{4}^{\infty}\de \mathcal{A}_p^{m}$) closely tracks the large-scale advection term. The two terms roughly balance each other, indicating a statistically steady state of advection and diffusion. This equilibrium does not imply that the poloidal magnetic field is static, but rather that any reconnecting or escaping field is constantly replenished by advection and the local regeneration of magnetic field.

Turbulent dissipation plays a different role for small-scale magnetic field loops than it does for a large-scale vertical magnetic field. For a large-scale vertical magnetic field, local dissipation in the disk can only diffuse the field towards larger radii \citep{lubow_magnetic_1994,guilet_transport_2012}. In contrast, for magnetic field loops smaller than the disk, dissipation acts as a local reconnecting agent. 

In the outer regions, $r>50$, the turbulent term $\sum_{4}^{\infty}\de \mathcal{A}_p^{m}$ is positive ; this sign change is attributed to the linear growth shown in Fig.~\ref{fig:energy_po}b. During the linear growth stage, before the local saturation time scale and a turbulent cascade develops, the "passive" modes generate magnetic energy. However, it is clear that the outer regions are not yet in a statistical steady state at this stage for $r>50$ (see Fig.~\ref{fig:Rad_beta}).

\begin{figure}
	\includegraphics[width=\columnwidth]{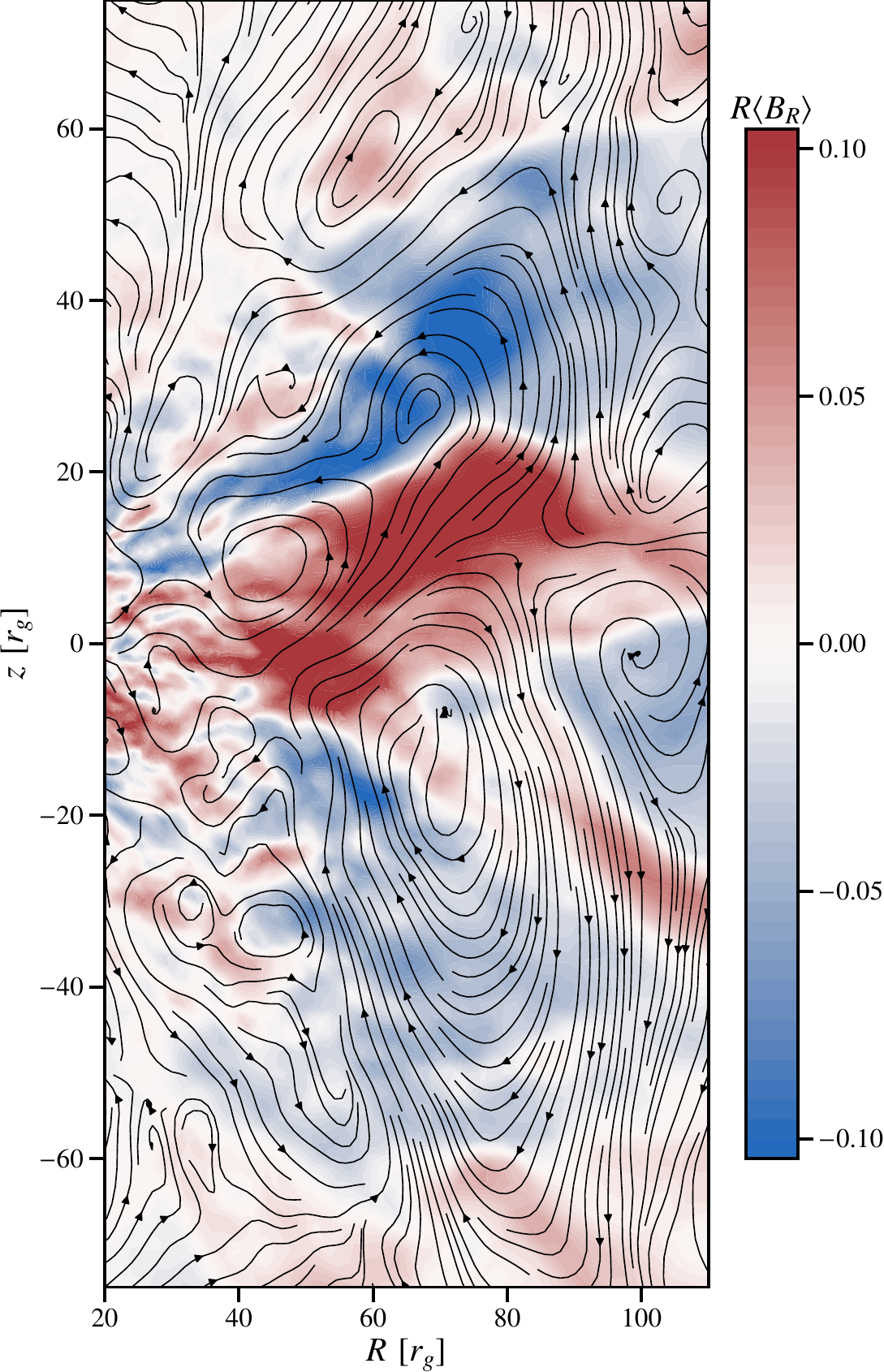}
    \caption{  Poloidal velocity streamlines, $\mathbf{u}_p^{(m=2)}$, of the $m=2$ mode at $\phi=\frac{4}{3}\pi$ superimposed on the axisymmetrized cylindrical radial field, $\mean{B_R}$. They take the form of rolling structures, whose centers are well correlated with the regions where $\mean{B_R}=0$. Poloidal magnetic-field generating motions thus takes the form of spiral rolling motions, which should not to be interpreted as axisymmetric eddies, as visualized here.}
    \label{fig:rollers}
\end{figure}

\subsection{Shearing wave structures}
As shown above, the dynamo mechanism relies on non-axisymmetric shearing wave structures, and it is thus interesting to analyze these structures in more detail. Figure \ref{fig:rollers} shows the poloidal velocity streamlines, $\mathbf{u}_p^{(m=2)}$, of the $m=2$ mode at $\phi=\frac{4}{3}\pi$ superimposed on the axisymmetrized cylindrical radial field, $\mean{B_R}$. Notice that the $m=2$ mode takes the form of rolling structures in the poloidal plane. These structures should not be interpreted as axisymmetric turbulent eddies; they are the rolling poloidal motions of traveling non-axisymmetric shearing waves. 
The centers of the rolling structures are well correlated with the places where the radial magnetic field changes sign. We interpret this rolling motion of the spiral modes as the field-generating action \citep{herault_periodic_2011}. Figure \ref{fig:rollers} should also be compared to Figure 7 of \cite{riols_magnetorotational_2017}, who found similar poloidal plane structures for the field-generating waves (although without the complexity of a global structure). They also found a correlation between the center of the wave structures and the radial magnetic field.

In the previous section, we estimated a regeneration time scale of the poloidal field as a function of radius, $t_{\rm g}=10\Omega_k^{-1}(R)$. To show the self-consistency of our analysis, we compute $t_{\rm g}$ using the amplitude of the wave velocities. We use the following estimate,
\begin{equation}
    u_{\rm sw} = \sum\limits_{m=1}^{m=3}\sqrt{\langle\left(\mathcal{R}(\mathbf{u}_p^{(m)})|_d\right)^{2}\rangle},
    \label{eq:usw}
\end{equation}
where $X|_d$ represents the disk average defined in Section \ref{sec:average}, $\mathcal{R}$ represents taking the real value. We square $\mathbf{u}_p^{(m)}$ before taking its azimuthal average because, $\langle \mathbf{u}_p^{(m)}\rangle =0$, due to it being a non-axisymmetric quantity.

\begin{figure}
	\includegraphics[width=\columnwidth]{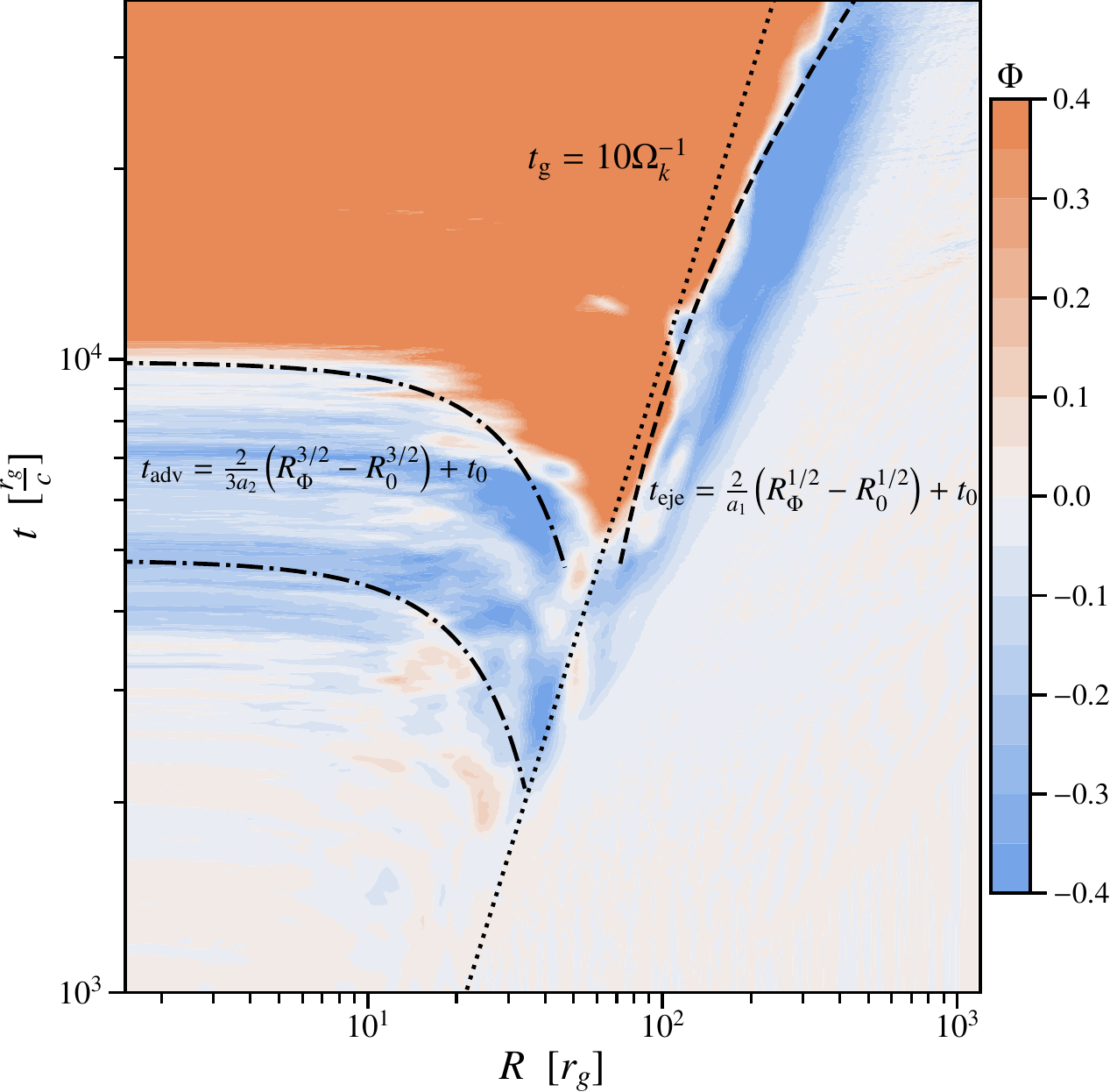}
    \caption{Axisymmetrized poloidal magnetic field flux (Eq.~\ref{Eq:pol_flux}), $\Phi$, evaluated at $\theta=\pi/2$, as a function of $t$ and $R$. We also show the growth time scale of the magnetic-field loop structures, $t_{\rm g}=10\Omega_k^{-1}$, (Eq.~\ref{Eq:growth_rate}) and the magnetic field advection and ejection trajectories, see text. The largest poloidal magnetic field structures appear along the growth time scale line, then clearly follow the advection and ejection trajectories.  }
    \label{fig:loops_advection}
\end{figure}

Figure \ref{fig:Radius_energyPo}b shows the estimate for the average wave amplitude, $u_{\rm sw}$, normalized to the local Keplerian velocity and time averaged between $t=5\times10^{3}r_g/c$ and $t=8\times10^{3}r_g/c$ as function of radius. We see that in the inner regions, that are in quasi steady state\footnote{Other estimates for $u_{\rm sw}$ lead to similar values, for example $u_{\rm sw} = |\mathcal{E}_\varphi|_d|/\mean{B_p}^2|_d\sim 10^{-2}V_K$. },
\begin{equation}
        u_{\rm sw} \simeq 2.3\times10^{-2}V_K(r).
\end{equation}
 If we assume that the wave structures generate magnetic fields on length scales the size of half the disk vertical thickness (see Fig.~\ref{fig:aphi1}a) we get
\begin{equation}
    t_{\rm sw} = \frac{h/2}{u_{\rm sw}} \sim 10\Omega_k^{-1}(r).
\end{equation}
This timescale is completely consistent with $t_{\rm g}$ computed in the previous section. This consistency is satisfying, as we can predict the field regeneration rate based on the amplitude of the waves. A computation using the vorticity, 
\begin{equation}
    \omega_{\rm sw} =  \frac{1}{2}\sum\limits_{m=1}^{m=3}\sqrt{\langle \rot\left(\mathcal{R}(\mathbf{u_p}^{(m)})\right)|_d^2\rangle}\simeq 0.2\Omega_k
\end{equation}
of the wave structures yields a similar time scale, $t_{\rm vort} = 1/\omega_{\rm sw}\simeq 20\Omega_k^{-1}$. However, this calculation does not show the role of the length scale in the dynamics.

By choosing $h/2$ as the length scale, we have illuminated a subtle issue. In Section \ref{sec:quali} we described that for one polarity to take hold, one magnetic field loop needs to be larger than the size of the disk scale height. Hence, if every magnetic field loop generated is of the size $h(r)/2$ there is no way for one polarity to take hold. A secondary mechanism is required to enhance the size of magnetic field structures. In Section \ref{sec:adv_size}, we argue that the advection serves this purpose.

\subsection{Field transport}
As demonstrated earlier, advection plays a crucial role in shaping the large-scale dynamics of the poloidal magnetic field by transporting magnetic flux from outer to inner regions. In this section, we provide a quantitative analysis of the velocities and time-scales associated with this field transport, elucidating their interaction with local magnetic field regeneration.

A simple afine fit for the radial velocity, gives
\begin{equation}
    \mean{u_r} = V_K(r)\left(a_1r-a_2\right),
    \label{Eq:fit_ur}
\end{equation}
where $a_1=7.8\times10^{-4}$ and $a_2=4.1\times10^{-2}$ (see Fig.~\ref{fig:Radius_energyPo}b). The value of $a_2$ is around a factor of $4$ larger than the value predicted by standard accretion theory\footnote{We note that this approximation ignores the radial gradient of $R^2\mean{P}$.}, $-\frac{\mean{u_r}}{V_k} \simeq \alpha_{\nu} \left(\frac{h}{r}\right)^2 \sim 10^{-2}$. We extrapolate Eq.~(\ref{Eq:fit_ur}) into two simpler limits for the radial velocity
\begin{align}
    \mean{u_r} \simeq a_1rV_K\,\,\,\mathrm{for}\,\,\,r\gg r_{\rm rev},\\
    \mean{u_r} \simeq -a_2V_K\,\,\,\mathrm{for}\,\,\,r\ll r_{\rm rev}.
\end{align}
Following \cite{jacquemin-ide_magnetic_2021} we use the expressions above to compute the trajectory of magnetic field lines. To do this we solve for the characteristics of field transport, given by $\diff {R}_\Phi/\diff t=\mean{u_r}$. This leads to the following trajectories

\begin{align}
    \label{eq:ejec_t}
     t_{\rm eje} &= \frac{2}{a_1}\left({R_\Phi^{1/2}}-R_0^{1/2}\right)+t_0\,\,\,\mathrm{for}\,\,\,r\gg r_{\rm rev},\\
    t_{\rm adv} &=\frac{2}{3a_2}\left(R_\Phi^{3/2}-R_0^{3/2}\right) +t_0 \,\,\,\mathrm{for}\,\,\,r\ll r_{\rm rev},
    \label{eq:adv_t}
\end{align}
where $t_{\rm adv}$ represents the poloidal field advection time scale and $t_{\rm ejec}$ is the poloidal field ejection time scale.
Using the expressions above, we can compute the transport of the large-scale loop that emerges as the dominant polarity at $t\sim 10^{4}r_g/c$. 
In Fig.~\ref{fig:loops_advection} we show an $r$-$t$ diagram for the poloidal magnetic flux (Eq.~\ref{Eq:pol_flux}), same as Fig.~\ref{fig:large_loops}. We show the regeneration timescale computed in Eq.~(\ref{Eq:growth_rate}) through Fig.~\ref{fig:Radius_energyPo}, $t_{\rm g}\simeq 10\Omega_k^{-1}$. The large-scale poloidal field structures appear along the line showing the regeneration timescale; $t_{\rm g}$ is a good estimate for the growth of large scale structures. After the axisymmetrized poloidal field loop is formed, it is advected either inward or outward.

We show the trajectories of an inward and outward advecting loops in Fig.~\ref{fig:loops_advection}. We compute the trajectory of the inward (outward) advecting loop for $R_0=50r_g$ ($R_0 = 70r_g$) and $t_0 = 4.5\times10^{3}r_g/c$. It is clear that our simple model, Eqs.~(\ref{eq:ejec_t}-\ref{eq:adv_t}), reproduces the large-scale advection of the magnetic field. Furthermore, magnetic field loops at earlier times, for $t<4\times10^{4}r_g/c$, follow the same kind of trajectories. We find that the flux is constantly advected at a velocity that follows a Keplerian scaling, in the inner regions, and a velocity that follows $\mean{u_r}\propto R^{1/2}$, in the outer regions. We conclude that the evolution of the axisymmetric poloidal flux and the emergence of the large-scale field are well described by the two timescales $t_{\rm g}$ and $t_{\rm adv}$.



\subsection{Advection and size of poloidal field structures}\label{sec:adv_size}
As the poloidal magnetic field is generated within the disk, the size of the magnetic field structures will be limited to the local disk scale height, $h(r)$. 
A simple way to build larger field structures than the local scale height without invoking a reverse cascade is provided by the large-scale transport, $\mathcal{A}_p$. Within the disk the poloidal magnetic field is advected towards the inner regions by the large-scale accretion flow, $\mean{u_r}$. Hence, if a loop is born at $r\simeq60\,\,r_g$ it needs to get advected to $r\simeq 30\,\, r_g$ to be larger than the disk thickness. We can see this happen in Fig.~\ref{fig:loops_advection}, a magnetic field loop generated at $r\simeq 60\,\,r_g$ slowly gets advected towards $r\simeq 20\,\,r_g$, and then experiences runaway growth by the ballooning instability \cite[see Section \ref{sec:quali} and][]{lynden-bell_why_2003}. A supplementary complication is that, if there are two or more loops of different polarities, they will compete over which can be advected and also thread the disk midplane faster. This competition will add some stochasticity to the evolution of the system as imbalances determine the dominant magnetic field polarity. The reason why the dominant polarity emerges from $r=60r_g$, instead of $r<60r_g$, is unclear, but we propose two complementary arguments:

(1) Magnetic field loops born before $t\sim 10\times(60 r_g)^{3/2}\sim 4.5\times10^{4}r_g/c$ do not have enough runway to be advected before being attached to the black hole (see fig.~\ref{fig:zoom_structure}). If a loop appears at $r = 30r_g$ it is larger than the disk scale height at $r\sim 15r_g$. However, at this distance the loop might already be too close to the ISCO and is thus doomed to connect to the BH before becoming larger. Thus there is a critical radius where the large-scale field is big enough to experience the ballooning instability without being trapped by the BH.

(2) We define the poloidal magnetic flux that the MRI dynamo can generate, $\Phi_{\rm loop}\simeq \mean{B_p}l_{\rm loop}^2$, where $l_{\rm loop}$ is the typical size of the loop. We have
\begin{equation}
    \mean{B_p}= \sqrt{8\pi\beta_p^{-1}}\sqrt{\mean{P}}\propto r^{-1},
\end{equation}
since $\mean{P}\propto r^{-2}$, and with $l_{\rm loop}\sim h/2$ we find that
\begin{equation}
    \Phi_{\rm loop} \simeq  \sqrt{8\pi\beta_p^{-1}}\sqrt{\mean{P}}\left(h/2\right)^2\propto r,
    \label{eq:aprox_flux}
\end{equation}
an increasing function of radius. Thus, the farther from the BH the generated loop is, the larger the magnetic flux it contains. While Eq.~(\ref{eq:aprox_flux}) is an approximation, we have verified its effectiveness in describing the time-averaged flux\footnote{Where the time average allows to reduce the contribution from advection}.
From this it is clear that there will be a critical radius, that depends on the initial magnetic field profile, where the magnetic flux generated will be enough to generate a large-scale field and saturate to a MAD. The previous argument implies that while a large-scale magnetic field (larger than the disk scale height) might be inevitable, the maximal magnetic flux that can be generated might depend on subtle details like disk scale-height, radial extent of the disk and initial magnetic field strength. The consequences of the maximal magnetic flux depending on disk geometry is discussed in section \ref{sec:discussion}.

For our simulation, the large-scale field then advects into the black hole until it reaches the MAD state \citep{tchekhovskoy_efficient_2011}. This advection phase can be seen in Figure 1 of \cite{liska_large-scale_2020}. {The dynamo mechanism is then probably stabilized by the large scale poloidal magnetic field \citep{salvesen_accretion_2016}.}

\section{Possible dynamo closures}\label{sec:clossure}
Mean-field models are useful tools to simplify the dynamics of complex systems by linearizing them. However, the MRI dynamo described in Section \ref{sec:main_dinam} is inherently nonlinear, so mean field models may be inadequate for accurately describing its dynamics.
Nonetheless, we attempt here to distinguish which of two common mean field models may be the most accurate for our nonlinear dynamics. In this section, we focus on a specific radii, but the trends described have been verified to be independent of this choice. 

The toroidal magnetic field is only generated by the $\Omega$-effect, while the turbulence only acts as a dissipating agent of it and could be modeled as a turbulent resistivity. In contrast, the poloidal magnetic field can be generated by turbulence and requires a more careful analysis.
As we showed in Section \ref{sec:main_dinam}, in statistical steady state only the large-scale MRI-unstable waves, $m=1,2,3$, generate magnetic fields, while smaller-scale modes dissipate the magnetic field. Thus, we only attempt to find a mean-field model for 
\begin{equation}
    \mathcal{E}_\varphi^{\rm sw}=\sum\limits^{3}_{m=1}\mathcal{E}_\varphi^{(m)}.
\end{equation}

\subsection{Alpha vs shear-current effects}
\begin{figure*}
	\includegraphics[width=0.85\textwidth]{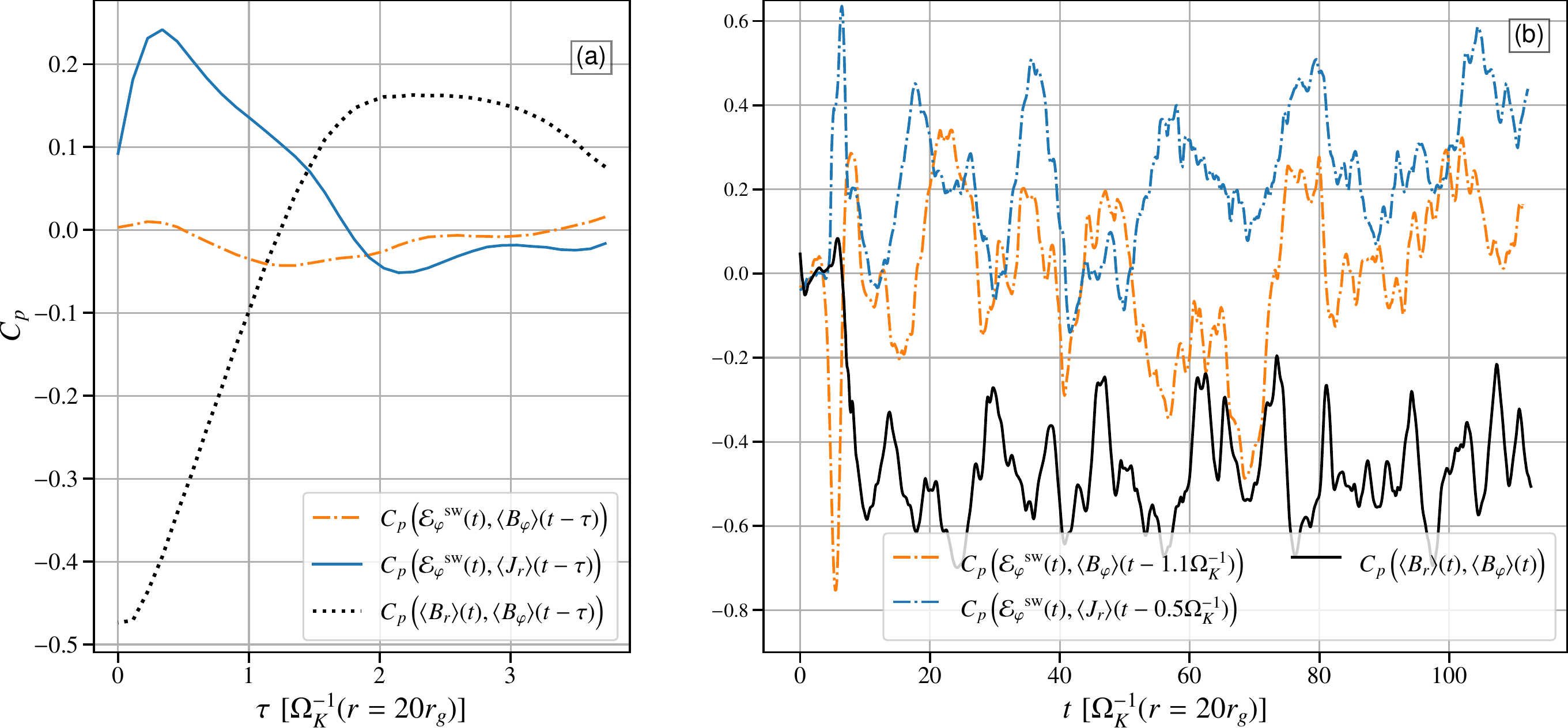}
    \caption{{\bf [Panel~(a)]:}Time and vertically averaged Pearson correlations at $r=20r_g$. The averages are computed between $t=20\Omega_K^{-1}(r=20)$ and $t=100\Omega_K^{-1}(r=20)$, and between $\theta_1 = \pi/2-\arctan(\frac{h}{R})$ and $\theta_2 = \pi/2-\arctan(0.1\frac{h}{R})$. The correlations are shown for both models, Eqs.~(\ref{eq:alpha_dyn}) and (\ref{eq:shear_cur}), and for the $\Omega$-effect which is used as a baseline of comparison. Only the $\alpha_d$-effect does not show a significant time average correlation. {\bf [Panel~(b)]:} Time variability of the Pearson correlation function, vertically averaged in the north hemisphere of the disk for $r=20r_g$. For every correlation we show the $t$ dependence at the $\tau$ maximum measured in panel~(a). Notice that on short time scales the $\alpha_d$-effect is comparable to the shear-current effect.}
    \label{fig:correl_timelag}
\end{figure*}

There are two main competing mechanisms for mean-field dynamo models of MRI turbulence. First, the classical alpha effect that has the following form,
\begin{equation}
    \mathcal{E}_\varphi^{\rm sw} = \alpha_{\rm d} \mean{B_\varphi},
    \label{eq:alpha_dyn}
\end{equation}
where $\alpha_{\rm d}$ is the alpha effect. This effect has been measured in different stratified shearing box simulations \citep{brandenburg_dynamo-generated_1995,gressel_mean-field_2010,gressel_characterizing_2015}
and global simulations \citep{flock_large-scale_2012,hogg_influence_2018,dhang_characterizing_2020}. Although all authors agree that the sign changes between the upper and lower hemispheres, there is no clear consensus on which sign should be measured on which hemisphere. 
The rough maximal magnitude measured in simulations is roughly consistent in the literature, $\alpha_{d}\sim\pm 5\times10^{-3}h\Omega_K$.

An alternative version of the $\alpha_d$ effect is the stochastic $\alpha_{\rm st}$ effect \citep{moffatt_magnetic_1978}. In this model, the mean $\overline{\alpha}_{\rm st} = 0$ but large fluctuations in $\sqrt{\overline{\alpha_{\rm st}^2}}$ can lead to magnetic-field growth \citep{brandenburg_magnetic_2008}, where $\overline{X}$ is a temporal average of $X$. This effect involves fluctuations in the mean field itself, which are thought to naturally occur when there is not a large-scale separation between the mean-field generating structures and the mean field \citep{brandenburg_advances_2018}.
\cite{heinemann_large-scale_2011} showed that a collection of shearing waves could excite a stochastic dynamo and generate large-scale magnetic field. Although the dynamo they computed is completely kinematic, and there is no backreaction of the magnetic field on the flow, the main features are reminiscent of the dynamics of our simulation. The fact that the dynamo is dominated by large-scale shearing waves might also explain its stochasticity. 

The shear current effect is another possible model for the MRI dynamo, it is expressed as
\begin{equation}
    \mathcal{E}_\varphi^{\rm sw} =- \eta_{\rm sc}\mean{J_r},
    \label{eq:shear_cur}
\end{equation}
where $\mean{J_r}$ is the electric current\footnote{We compute $\vmean{J}$ in its Newtonian approximation, neglecting the displacement current, a valid approximation at $r=20r_g$.}. This effect was identified analytically \citep{lesur_localized_2008} and also measured in non-stratified shearing boxes \citep{lesur_self-sustained_2008,squire_generation_2015,squire_magnetic_2016}. Crucially $\eta_{\rm sc}<0$ is a necessary condition for this dynamo to operate \citep{squire_generation_2015,squire_magnetic_2016,rincon_dynamo_2019}.

\citet{lesur_self-sustained_2008} found that the shear current effect requires a time delay, $\tau$, to model the dynamo cycle. This time delay is understood as the time it takes for the mean field to excite waves that will amplify the magnetic field, they assume $\tau \sim \Omega_k^{-1}$.

To distinguish which model best fits our nonlinear dynamo, we introduce the following time-delayed moving-averaged Pearson correlation function:

\begin{equation}
\label{eq:cp}
    C_{\rm p}(X,Y(t-\tau)) = \frac{\int\limits^{t}_{t-\delta t} X(t')Y(t'-\tau)\diff t'}{\sqrt{\int\limits^{t}_{t-\delta t} X^2(t')\diff t'\int\limits^{t}_{t-\delta t} Y^2(t'-\tau)\diff t'}},
\end{equation}
where $\tau$ is the time delay and $\delta t = n\Omega_K+\tau$ is the averaging window\footnote{The inclusion of $\tau$ in $\delta t$ is to ensure that our average window is always larger than the time shift.}. We choose $n=4$ to ensure that we average over a few Keplerian shear time scales, we find no perceptible differences for different values of $n$ as long as $1<n<10$. 
\begin{figure}
	\includegraphics[width=\columnwidth]{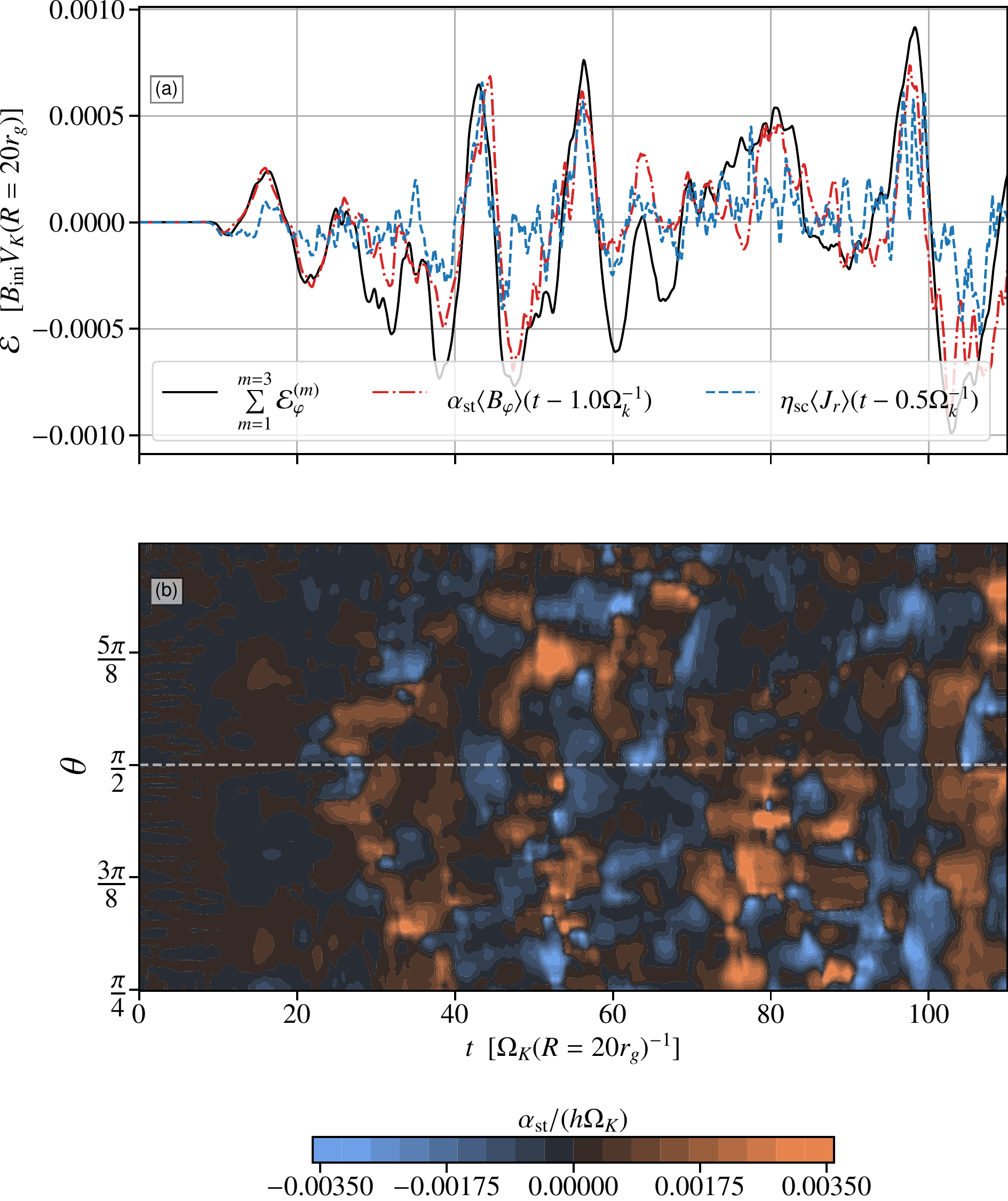}
    \caption{{\bf [Panel (a)]:} Time evolution of the full $\mathcal{E}_\varphi^{\rm sw}$, compared to mean-field EMFs using $\alpha_{\rm st}$ and $\eta_{\rm sc}$; coefficients computed from Eq.~(\ref{eq:fit_alpha}) and Eq.~(\ref{eq:fit_eta}), respectively. We have latitudinally averaged $\mathcal{E}_\varphi^{\rm sw}$ and its models over the northern hemisphere of the disk at $r=20r_g$. Both model roughly reproduce the EMF with a slight edge for the stochastic alpha model. {\bf [Panel (b)]:} $\theta-t$ diagram of the stochastic alpha, $\alpha_{\rm st}$, computed using Eq.~(\ref{eq:fit_alpha}) normalized to $h\Omega_k$. The  coefficient $\alpha_{\rm st}$ is temporally incoherent and lacks the anti-symmetry observed in most stratified shearing box simulations.}
    \label{fig:Fit_EMF}
\end{figure}

We show the time-averaged Pearson correlations of Eq.~(\ref{eq:cp}) in \autoref{fig:correl_timelag}a, between $t=20\Omega_K^{-1}(r=20r_g)$ and $t=100\Omega_K^{-1}(r=20r_g)$, for $r=20r_g$. The correlations are also averaged in the northern hemisphere of the disk, between $\theta_1 = \pi/2-\arctan(\frac{h}{R})$ and $\theta_2 = \pi/2-\arctan(0.1\frac{h}{R})$. We first compute the correlation function for all $\theta$ and then average it. We show the correlations of both models Eqs.~(\ref{eq:alpha_dyn}) and (\ref{eq:shear_cur}) and the correlation of $\mean{B_r}$ and $\mean{B_\varphi}$, which we use as a baseline of comparison.

We see that the correlation of the radial and toroidal fields peaks at $\tau=0$ with a maximal value of $-0.5$. The position of the peak at $\tau=0$ is consistent with non-lagged shear, which is what is expected from the induction equation, as the $\Omega$-effect does not include any lag. However, the maximal value of only $-0.5$ shows how difficult it is to find correlations in global GRMHD simulations. Other global GRMHD simulations find clean correlations of the order of $0.3$, although for neutron star dynamos \citep{kiuchi_large-scale_2023}.
\begin{figure}
	\includegraphics[width=\columnwidth]{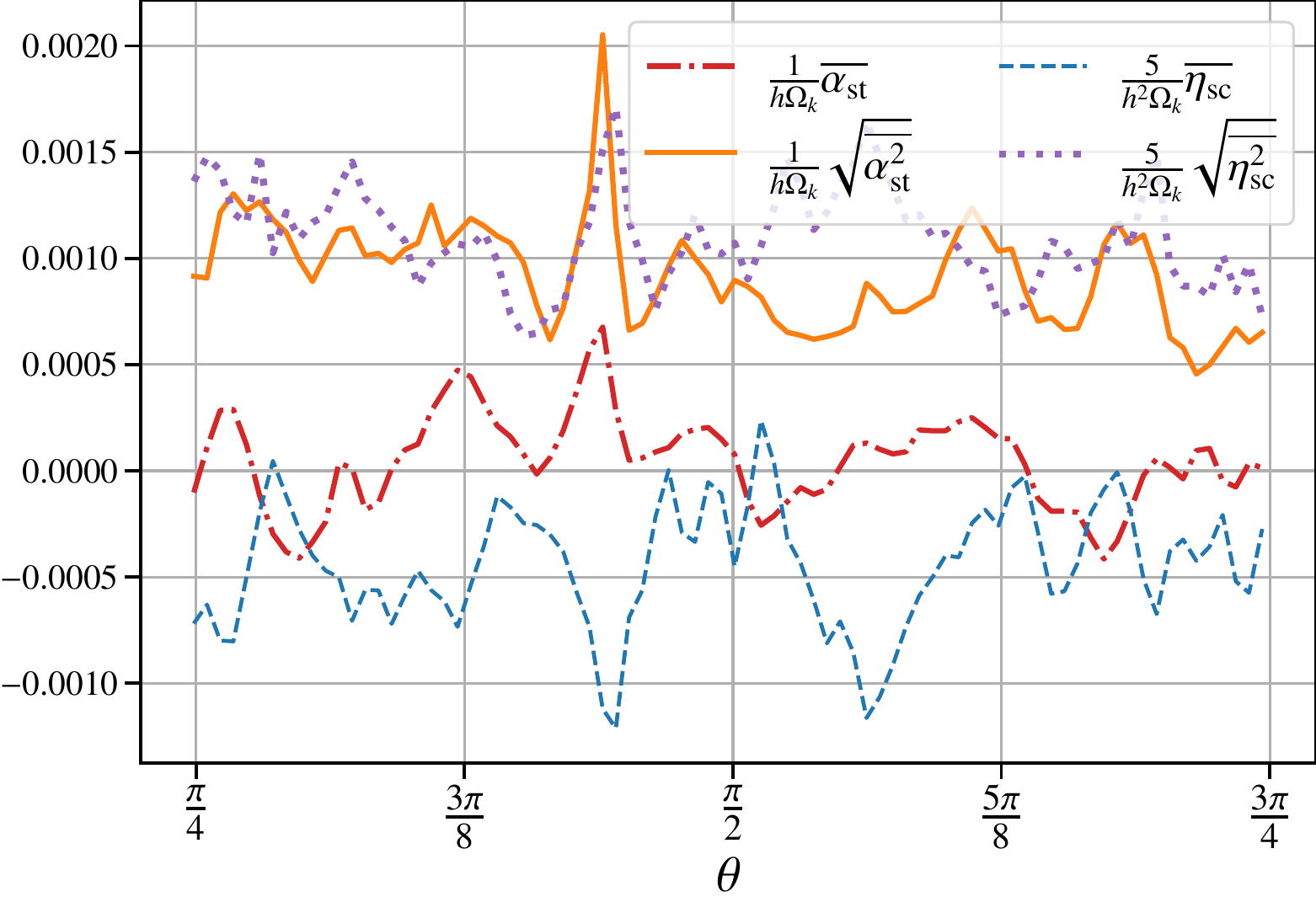}
    \caption{Time-averaged $\overline{\alpha_{\rm st}}$ and $\overline{\eta_{\rm sc}}$ coefficients as functions of the polar coordinate, normalized to $h\Omega_k$ and $h^2\Omega_k$ respectively and evaluated at $r=20r_g$. We also show their respective rms values $\sqrt{\overline{\alpha_{\rm st}^2}}$ and $\sqrt{\overline{\eta_{\rm sc}^2}}$. For the sake of visibility the shear current coefficients are multiplied by $5$.
    The coefficients are computed using Eq.~(\ref{eq:fit_alpha}) and Eq.~(\ref{eq:fit_eta}), respectively. No clear trend is observed in the polar profile of $\alpha_{\rm st}$ and its rms value is about a factor of $2$ larger. The shear-current prescription, $\eta_{\rm sc}$, shows a clearer trend hitting a maximum just above the disk midplane.}
    \label{fig:coeffs_fits}
\end{figure}

The correlation for a shear-current-effect-like precription (Eq.~\ref{eq:shear_cur}) shows a peak at $\tau=0.5\Omega_K^{-1}(r=20r_g)$, consistent with \citet{lesur_self-sustained_2008}. In other words, the $\mathcal{E}_\varphi^{\rm sw}$ reacts to the radial current with a $\tau$ of the order of the MRI growth rate. The maximal value of $C_p(\mathcal{E_\varphi}(t),\mean{J_r}(t-\tau))$ is $0.25$, a marginally significant correlation. However, the correlation for an $\alpha_d$ effect prescription (Eq.~\ref{eq:shear_cur}) is far smaller, showing a maximum of only $-0.05$, and it also shows a maximum at a time lag of $\tau = \Omega_K^{-1}(r=20r_g)$. 

Figure \ref{fig:correl_timelag}b shows the time variability of the Pearson correlation functions, vertically averaged in the north hemisphere of the disk for $r=20r_g$ for their respective $\tau$ maximum measured in Fig.~\ref{fig:correl_timelag}a:
\begin{itemize}
    \item $\tau =0$ for $C_p(\mean{B_\varphi}(t),\mean{B_\varphi}(t-\tau))$
    \item $\tau = 0.5\Omega_K^{-1}(r=20r_g)$ for $C_p(\mathcal{E_\varphi}^{\rm sw}(t),\mean{J_r}(t-\tau))$
    \item $\tau = 1.1\Omega_K^{-1}(r=20r_g)$ for $C_p(\mathcal{E_\varphi}^{\rm sw}(t),\mean{B_\varphi}(t-\tau))$
\end{itemize}
All correlations are large, with $C_p>0.4$, and show strong variability. We notice that even though the average correlation between the mean toroidal field $\mean{B_\varphi}$ and the electromotive force $\mathcal{E_\varphi}^{\rm sw}$, $C_p(\mathcal{E_\varphi}(t),\mean{B_\varphi}(t-\tau))$, is small, it can reach high correlation on shorter time scales. This large deviation from the average could be a sign of a stochastic $\alpha_{st}$ dynamo effect. We find identical trends for the northern hemisphere of the disk. For the sake of concisenes, we show the $t$-$\tau$ diagram in Appendix \ref{A:timelag}.

\subsection{Mean-field EMFs}

For large enough correlations, it makes sense to approximate the $\alpha_{\rm st}$ and $\eta_{\rm sc}$ coefficient with
\begin{align}
    \label{eq:fit_alpha}
    \alpha_{\rm st}&\simeq \frac{\int\limits^{t}_{t-\delta t} \mathcal{E}^{\rm sw}_\varphi(t')\mean{B_\varphi}(t'-\tau)\diff t'}{\int\limits^{t}_{t-\delta t} \mean{B_\varphi}^2(t')\diff t'},\\
    \label{eq:fit_eta}
    \eta_{\rm sc}&\simeq \frac{\int\limits^{t}_{t-\delta t} \mathcal{E}^{\rm sw}_\varphi(t')\mean{J_r}(t'-\tau)\diff t'}{\int\limits^{t}_{t-\delta t} \mean{J_r}^2(t')\diff t'}.
\end{align}
Due to variability of $C_{\rm p}$, this equation only makes sense for small $\delta t>\tau$. We choose $\delta t =5\Omega_k^{-1}(r=20r_g)$. We use the same values of $\tau$ as the ones described above (see also Fig.~\ref{fig:correl_timelag}b).

Figure \ref{fig:Fit_EMF}a shows the time evolution of the full $\mathcal{E}_\varphi^{\rm sw}$, compared to the two different models. With $\alpha_{\rm st}$ and $\eta_{\rm sc}$ coefficients computed from Eq.~(\ref{eq:fit_alpha}) and Eq.~(\ref{eq:fit_eta}), respectively. 
Both models roughly reproduce $\mathcal{E}_\varphi^{\rm sw}$, with a noticeable edge for the $\alpha_{\rm st}$ model. We tried different values of $\delta t$: the fit deteriorates for larger values of $\delta t$, but the $\alpha_{\rm st}$ model is always a marginally better fit to the full $\mathcal{E}_\varphi^{\rm sw}$.

Figure \ref{fig:Fit_EMF}b shows the $\theta$-time diagram of $\alpha_{\rm st}$ evaluated at $r=20r_g$. As expected, the alpha coefficient is highly variable and does not show a clear top down symmetry, in contrast with other works \citep{flock_large-scale_2012,hogg_influence_2018}. We do not show the temporal evolution $\eta_{\rm sc}$ as it shows similar variability to the stochastic $\alpha_{\rm st}$. 

Instead, we show in Fig.~\ref{fig:coeffs_fits} the temporal average ($\overline{\alpha_{\rm st}}$, $\overline{\eta_{\rm sc}}$) and rms value ($\sqrt{\overline{\alpha_{\rm st}^2}}$, $\sqrt{\overline{\eta_{\rm sc}^2}}$) of both coefficients, computed between $t_1=0$ and $t_2=110\Omega_k^{-1}$. We find no clear trends for the polar profile of $\overline{\alpha_{\rm st}}$, although there might be a slight change of sign around the disk midplane. This signal is likely overwhelmed by the large fluctuations, as the rms value is around a factor of $5$ larger.
We compute a rms value for the stochastic $\alpha$ prescription of $\sqrt{\overline{\alpha_{\rm st}^2}}\simeq 10^{-3}h\Omega_k$, which is roughly constant in $\theta$.

As far as the shear-current effect is concerned, we compute a clearly negative shear-current coefficient $\eta_{\rm sc}\simeq-2\times10^{-4}h\Omega_k^2$ at its maximal amplitude; the sign of the $\eta_{\rm sc}$ coefficient thus allows for magnetic field amplification \citep{rincon_dynamo_2019}.
In contrast with $\overline{\alpha_{\rm st}}$, the shear current effect shows a slight trend in $\theta$, reaching its maximum slightly above the disk midplane. Furthermore, in this case the maximal value is of the same order as the rms fluctuations.

From the previous analysis, tt is difficult to conclude which model is the better fit for the MRI dynamo.  While the stochastic alpha model better fits the behavior of the EMF when compared the shear current effect, it has a worse average correlation. However, the stochastic alpha model is more intuitively appealing as it is reminiscent of the wave phenomenology detailed in Section \ref{sec:main_dinam} \citep{heinemann_large-scale_2011}.

The other main conclusion of this mean-field analysis is that the antisymmetric (with respect to the disk midplane) alpha coefficient computed in stratified shearing box simulations is incompatible with our 3D global simulations \citep{brandenburg_dynamo-generated_1995,gressel_mean-field_2010}. There are two possible explanations for this discrepancy: (1) Stratified shearing boxes are known to have too many symmetries \citep{lesur_magnetorotational_2013,lesur_magnetohydrodynamics_2020}. The additional symmetries of shearing boxes could lead to the organization of the large-scale wave structures, which could lead to a more coherent $\alpha$ effect.  Furthermore, both global simulations where a coherent alpha effect was measured had a limited latitudinal extent \citep{flock_large-scale_2012,hogg_influence_2018}. This limited latitudinal extent could lead to a similar effect of organizing the wave structures. (2) A thinner disk could also lead to a more coherent $\alpha$ effect by vertically confining the wave structures. Indeed, \cite{hogg_influence_2018} found that thicker disks had more irregular dynamo cycles. Future work will focus on how this dynamo changes in thinner disks.

Finally, \cite{dhang_characterizing_2020} used singular value decomposition to measure mean field dynamo coefficients in RIAFs. Although they use a different initial field, the overall structure looks similar to ours before $t=8\times10^{3}r_g/c$. They were only able to fit $\alpha_d$ coefficient at the surface of the disk, and not within the interior of the disk, and attributed this lack of convergence to the disk being too turbulent. This observation might be related to the intrinsic variability that we show in Fig.~\ref{fig:Fit_EMF}b.

\section{Conclusions \& Discussion}\label{sec:conclu_discu}
\subsection{Main conclusions}
Binary mergers are theorized to drive jets as a component of the GRB and afterglow emission. If the central engine is a BH, the jet will be driven by the \cite{blandford_electromagnetic_1977} process. However, the BZ mechanism requires a large-scale poloidal magnetic field.
In this paper, we analyzed simulations of GRMHD accretion disk to investigate the dynamical mechanism underlying the generation, transport and self-organization of the magnetic field.  We have here isolated a mechanism that can generate the large-scale poloidal magnetic fields required for jet launching. 

We elucidated that, excluding large-scale transport, the production and dissipation of magnetic energy relies on the following ingredients: (1) The toroidal magnetic field is generated through the shear of the poloidal field (Fig.~\ref{fig:To_energy}). (2) Large-scale non-axisymmetric MRI wave structures of wavenumber, $m=1,2,3$, nonlinearly generate the poloidal field (Fig.~\ref{fig:energy_po}). (3) The toroidal field loses energy to those non-axisymmetric field fluctuations (Fig.~\ref{fig:To_energy} and Appendix \ref{A:multiwaves}). (4) Finally, both field components lose energy to the small-scale, $m>3$, non-axisymmetric structures through turbulent dissipation. 
This behavior is highly reminiscent of previous results obtained in simpler shearing box simulations \citep{lesur_self-sustained_2008,herault_periodic_2011,riols_dissipative_2015,riols_magnetorotational_2017}.

As described in the introduction, this dynamo mechanism is interpreted as an interplay between the non-axisymetric MRI and large-scale axisymmetric dynamics of the magnetorotational instability. The low-$m$ modes are (non-)linear non-axisymmetric MRI modes that feed on the shear and develop on the axisymmetric toroidal magnetic field. The low $m$-modes take the form of MRI-amplified shearing waves (see Fig.~\ref{fig:rollers}), which are sheared spiral waves \citep{goldreich_ii_1965,johnson_magnetohydrodynamic_2007}. In the statistical steady state, the large-$m$ modes ($m>4$) are non-linearly generated by the turbulent cascade and mix the large-scale field, dissipating it all small scales. We computed a local growth time scale for the magnetic-field-generation mechanism and found $t_{\rm g}\simeq 10\Omega_k^{-1}$. 
The velocity of the spiral modes, $u_{\rm sw}$, can the be used to predict this saturation timescale, $t_{\rm g}\sim t_{\rm sw}=h/u_{\rm sw}$, with good precision (see Fig.~\ref{fig:Radius_energyPo}b).
Hence, magnetic field loops are generated locally at a specific radii with a frequency $10$ times slower than the Keplerian frequency. 

We also found that large-scale advection of the magnetic field, poloidal or toroidal, is an essential feature of the full magnetic-field dynamics. We showed that the toroidal magnetic field vertically escapes the accretion disk (Fig.~\ref{fig:transport_tor} and Fig.~\ref{fig:eneg_avgTo}). 
Most importantly, the poloidal magnetic field is advected inwards by the accretion flow. 
The advection serves as a scale amplification mechanism by transporting the larger-scale loops, generated in the outer regions of the accretion disk, towards the inner regions, with a smaller disk scale height. Thus, advection can transport a magnetic field larger than the local disk scale height. We computed the advection timescale, $t_{\rm adv}$,  over which this happens using a simple fit for the accretion flow velocity (see Eq.~\ref{eq:adv_t}). We have determined that this straightforward calculation of the advection timescale is remarkably consistent with magnetic field transport and, as a result, possesses predictive capabilities. When putting everything together we can formulate a description for the emergence of a dominant large-scale vertical magnetic field using only two timescales, $t_{\rm g}$ and $t_{\rm adv}$. We have also approximated the magnitude and profile of the magnetic flux and found that it depends mostly on the geometry of the disk.

Finally, even though the dynamo described here is completely nonlinear, for the purpose of global modeling and better closures, we attempted to determine which linear mean-field model is capable of reproducing the main features of our simulations. We found that a simple antisymmetric $\alpha$ dynamo model, often computed in stratified shearing box simulations, can not reproduce the dynamo mechanism observed in our simulation. However, a shear current effect or a stochastic $\alpha$ dynamo could reproduce some of the main features. We believe this discrepancy with shearing box simulations is due to the thickness of our accretion disk. Thinner disk, which are well modeled by shearing boxes, might confine and organize the field-generating structures, and the organization of such structures might lead to a more coherent $\alpha$ effect.

\subsection{Discussion}\label{sec:discussion}

\subsubsection{Dependence on initial field geometry and strength}
In this work, we have only considered an initially axisymmetric toroidal magnetic field. It is not entirely clear how sensitive our results are to these initial conditions. However, as noted earlier in Section \ref{sec:setup}, an approximately axisymmetric toroidal field seems to emerge in the initial phase of evolution of BHNS mergers, $t<100\,\,\mathrm{ms}$ \citep{aguilera-miret_role_2023}. This is encouraging, as this axisymmetric toroidal field plays a key meadiating role in MRI mechanism described in this work. Furthermore, \cite{gottlieb_large-scale_2023} found that their non-axisymmetric initial condition, consisting of two large-scale toroidal field features with opposite polarities, still led to the generation of large-scale poloidal fields. The authors have verified that the main features of the mechanism described in this manuscript are also present in the simulation of \cite{gottlieb_large-scale_2023} (private communication). This suggests that the MRI dynamo described in this work is robust to non-axisymmetric initial conditions. 

We also found no difference in the magnetic field generation mechanism between a net large scale toroidal field and an anti-symmetric, zero net flux toroidal field (Appendix \ref{A:sim2}). This independence on the net toroidal flux is encouraging, as it could be that the mechanism can be excited with smaler scale toroidal field structures.

\cite{dhang_magnetic_2023} found that for an initial weak, $\beta_{\rm ini}=800$, zero net flux poloidal field, the accretion disk was unable to reach the MAD state after $120000r_g/c$. Although no MAD was present, a coherent large-scale magnetic field appeared, as shown in their Figure 7 and the last panels of Figure 2. However, this field did not have enough flux to reach the MAD state. Similarly, \cite{rodman_magnetic_2023} recently found that initially weak toroidal magnetic field, $\beta_{\rm ini}=200$, would not enter the MAD state after $120000r_g/c$. Their results show that weaker magnetic fields might substantially weaken the efficiency of the nonlinear magnetic field generation mechanism shown here.

However, in both studies the authors either imposed a null field for $r>200r_g$ or considered a limited radial domain size $r_{\rm max}<300r_g$. As was argued on Section \ref{sec:adv_size} the maximal magnetic flux that can be generated is set by the size of the accretion disk, including its radial extent. It would be interesting to consider more weakly magnetized disks like the ones simulated by \cite{rodman_magnetic_2023} and \cite{dhang_magnetic_2023}, but with a larger radial extent. Following the arguments of Section \ref{sec:adv_size}, this could lead to the formation of a MAD. \cite{rodman_magnetic_2023} also ran a simulation with a stronger initial magnetic field, $\beta_{\rm ini}=5$, that reached a state more reminiscent of a MAD and might be qualitatively similar to the simulation presented here. 

\subsubsection{Reaching the MAD state}
An intriguing implication of our study is that the maximum magnetic flux generated is determined by the radial extent and geometrical thickness of the accretion disk \citep[see for example][where different final steady states are computed]{dhang_magnetic_2023,rodman_magnetic_2023}. This constraint on the maximal magnetic flux that can be generated is specially relevant for mergers, where the radial extent of the disk is limited. At first glance this might be bad news for the MAD state. However, it should be noted that the transition to this MAD state also depends on the accretion rate. For a finite mass reservoir, $\dot{m}$ will drop, and the MAD state will be eventually reached regardless of the maximal magnetic flux that can be generated \citep{tchekhovskoy_magnetic_2015,tchekhovskoy_launching_2015,gottlieb_large-scale_2023}. However, it is crucial to note that if the system attains the MAD state toward the end of its existence, then the MAD state becomes irrelevant to the system's dynamics.

The study by \cite{gottlieb_unified_2023} suggests that the X-ray power-law decay observed in the extended emission of compact merger GRBs may be attributed to the system transitioning to the MAD state. If this hypothesis holds, the maximal poloidal magnetic flux that can be generated becomes the primary controlling parameter, along with the initial mass of the disk, for the duration of the GRB event.
An exploration of the potential maximal magnetic fluxes based on various parameters of the binary merger, such as the central engine, disk mass and size, could be approximated using our work as a foundation (see Section \ref{sec:adv_size}). This analysis is deferred to future investigations but has the potential to provide complementary insights into understanding the unified evolution of merger GRBs \citep{gottlieb_unified_2023}. This analysis could also be useful for models of TDEs \citep{teboul_unified_2023}. In such models, the jet efficiency is free parameter, one could use our prescriptions for the generation and transport of magnetic flux to self-consistent compute the jet efficiency.



Finally, the dynamo mechanism outlined in this study, relying heavily on advection, may encounter challenges when the accretion disk becomes geometrically thin. Thinner disks are notably less proficient in the advection of magnetic fields \citep{lubow_magnetic_1994,jacquemin-ide_magnetic_2021}. It is a well-established fact that thin disks, such as those found during the soft state in X-ray binaries \citep{done_modelling_2007}, exhibit reduced tendencies to launch relativistic jets. Consequently, it becomes imperative to investigate the mechanism's dependence on the disk's thickness, as this insight may shed light on why thinner disks tend to be less predisposed to jet launching.

\subsubsection{Stochasticity and flares}

Due to the stochastic nature of the dynamo mechanism, simulations initially set with toroidal magnetic fields exhibit an initial bursty behavior in jet power and efficiency, preceding the emergence of a dominant polarity \citep[see Fig.~\ref{fig:loops_advection} and][]{christie_role_2019,gottlieb_large-scale_2023}. This early activity mirrors the quasi-periodic precursor flares observed in short Kilonova gamma-ray bursts (GRBs) \citep{xiao_quasi-periodically_2022}. Our findings indicate that this behavior is attributed to the early attachment of a magnetic loop to the black hole (see Fig.~\ref{fig:zoom_structure}). Although the topology of this attached loop differs from that observed in magnetically arrested disks (MADs), it still facilitates the launching of a Blandford-Znajek (BZ) jet. It is worth noting that the jet produced by this small-scale loop is less efficient compared to those driven by MADs \citep{christie_role_2019}. The cancellations of magnetic flux on the black hole event horizon occur as additional magnetic loops of opposite polarities are advected, like the simulations of \cite{parfrey_black_2015}. These cancellations give rise to oscillations in magnetic flux, which may explain the quasi-periodic precursor flares observed in short GRBs. Such variability could also explain the initially bursty X-ray luminosity curve of the TDE Swift J1644+57, that is normally modeled with precession of the accretion disk \citep{tchekhovskoy_swift_2014,teboul_unified_2023}.

As time progresses, a truly large-scale magnetic field is generated and the MAD state is reached. This large-scale magnetic flux may be responsible for the extended emission observed for $t>1\rm{s}$ in short GRBs \citep{norris_short_2006,gottlieb_unified_2023}. Further investigation of this scenario is warranted, particularly with simulations initialized with weaker magnetic fields.



\cite{el_mellah_spinning_2022} examined a hybrid magnetic topology, with closed and open field lines connected to the BH, as shown in Figure \ref{fig:zoom_structure}. They used 2D GR particle-in-cell simulations to find that this hybrid configuration could drive efficient jets and particle acceleration through reconnection sheets. 
In this work, we showed that such a configuration naturally emerges in the early stages of the evolution of the MRI dynamo. This provides a natural justification for their \emph{previously ad-hoc} configuration.

Overall, the work presented here, anchored in a detailed analysis of 3D GRMHD dynamics, provides new insights into the physical mechanism underlying the global magnetized dynamics and magnetic field generation in black-hole accretion. In particular, simulations of mergers are usually initialized with artificially large magnetic fields due to constrains on resolving the magnetized turbulence. While subgrid models might alleviate some of the issues related to resolution, they rely on assumptions and fine tuned coefficients. Understanding the complicated nonlinear dynamics therefore paves the way towards better mean-field models that are complementary to the subgrid approaches. Overall, the work presented here provides a contribution to better global large-scale models of BH accretion





\section*{Acknowledgements}
JJ thanks Geoffroy Lesur and Jonathan Ferreira for their helpful feedback on the results and Antoine Riols, Nick Kaaz, and Beverly Lowell for valuable discussions.

JJ and AT acknowledge support by the NSF AST-2009884 and NASA 80NSSC21K1746 grants. AT also acknowledges support by NSF grants AST-2107839, AST-1815304, AST-1911080, OAC-2031997, and AST-2206471. Support for this work was also provided by the National Aeronautics and Space Administration through Chandra Award Number TM1-22005X issued by the Chandra X-ray Center, which is operated by the Smithsonian Astrophysical Observatory for and on behalf of the National Aeronautics Space Administration under contract NAS8-03060. This research was also made possible by NSF PRAC award no. 1615281 at the Blue Waters sustained-petascale computing project and supported in part under grant no. NSF PHY-1125915. This research used resources of the Oak Ridge Leadership Computing Facility, which is a DOE Office of Science User Facility supported under Contract DE-AC05-00OR22725 via ALCC, INCITE, and Director Discretionary allocations PHY129. ML was supported by the John Harvard, ITC, and NASA Hubble Fellowship Program fellowships.




\bibliographystyle{mnras}
\bibliography{example} 




\appendix

\section{Energy generation or dissipation of the first eight non-axisymmetric modes}\label{A:multiwaves}
In this appendix, we confirm that the first three non-zero $m$-modes, referred to as active modes, are the sole contributors to the generation of poloidal magnetic energy. Furthermore, we show that these modes dissipate toroidal magnetic energy and facilitate angular momentum transport. In essence, the active modes are identified as MRI-driven shearing waves, mediated by the axisymmetric toroidal field. 

\subsection{Poloidal magnetic energy}
\begin{figure}
	\includegraphics[width=0.95\columnwidth]{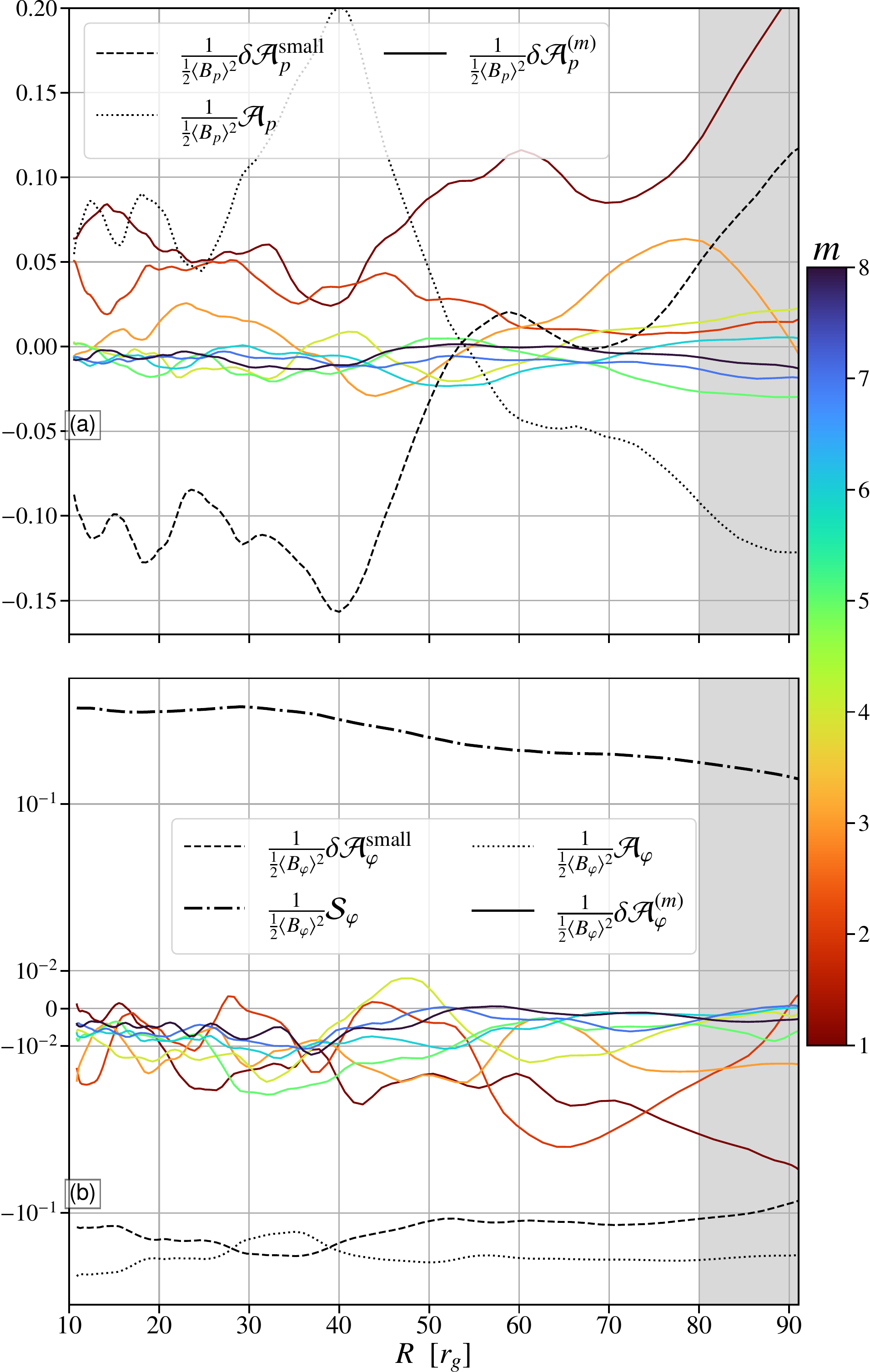}
    \caption{ Vertically and temporally averaged poloidal (a)  (Eq. \ref{Eq:pol_energ}), and toroidal (Eq.~\ref{Eq:tor_ener}) (b) energy equation normalized to the local Keplerian frequency. The temporal average is performed between $5000r_g/c$ and $8000r_g/c$. The different terms are divided by the vertical and temporal average of $\mean{B_p}^2$ (a) or $\mean{B_\varphi}$ (b). (a) The colored solid lines show the first 8 non-axisymetric $m$-modes, $\delta\mathcal{A}_p^{(m)}$ (a) or $\delta\mathcal{A}_\varphi^{(m)}$ (b). The dotted line shows the large-scale axisymmetric transport of poloidal (a) or toroidal (b) field. The dashed line shows the small-scale turbulent transport without the first eight non-axisymetric modes, $ \delta \mathcal{A}_p^{\rm small}$ (a) or $ \delta \mathcal{A}_\varphi^{\rm small}$. The dotted-dashed line shows the shear of poloidal into toroidal field, $\mathcal{S}_\varphi$. }
    \label{fig:multiwav}
\end{figure}

In Fig.\ref{fig:multiwav}a, we present the various source terms for poloidal magnetic energy (Eq.~\ref{Eq:pol_energ}). These terms are vertically averaged within the disk, temporally averaged between $5000\,\,r_g/c$ and $8000\,\,r_g/c$, and normalized by the averaged axisymmetric poloidal magnetic energy, $\frac{1}{2}\mean{B_p}^2$. The sources of poloidal magnetic energy are categorized into three components: (1) The contributions of the first eight non-axisymmetric $m$-modes, denoted as $\delta\mathcal{A}_p^{(m)}$, as defined in Eqs.~(\ref{eq:Ap_mode}) and (\ref{eq:EMF_mode}). (2) The large-scale axisymmetric transport of poloidal field, represented by $\mathcal{A}_p$. (3) The small-scale turbulent transport, excluding the first eight non-axisymmetric modes,
\begin{equation}
    \delta \mathcal{A}_p^{\rm small} = \delta \mathcal{A}_p - \sum\limits_{m=1}^{8}\delta\mathcal{A}_p^{(m)}.
\end{equation}
Figure \ref{fig:multiwav}a confirms that the only modes that generate axisymmetric poloidal magnetic energy on average are $m=1,2,3$, where $m=1,2$ are the main contributors. We notice that $\delta \mathcal{A}_p^{\rm res} \simeq \sum\limits_{m=3}^{\infty}\delta\mathcal{A}_p^{(m)}$, as this term is dominated by the more numerous large $m$ modes, see Fig.~\ref{fig:Radius_energyPo}. Consequently, we refrain from providing a detailed analysis of the magnitude and behavior of $\mathcal{A}_p$ and $\delta \mathcal{A}_p^{\rm res}$, as this information is discussed in Section \ref{sec:main_dinam}.

\subsection{Toroidal magnetic energy}
We analyze the sources of toroidal magnetic energy by decomposing them into their non-axisymmetric $m$-modes. Extending the same $m$-mode decomposition applied to the poloidal magnetic field, we obtain
\begin{equation}
    \delta \mathcal{A}_\varphi = \sum\limits_{m=1}^\infty\de\mathcal{A}_\varphi^{(m)},
    \label{eq:parseval_phi}
\end{equation}
where
\begin{equation}
    \de\mathcal{A}_\varphi^{(m)} = \vmean{B_\varphi}\mathbf{e_\varphi}\cdot\rot\left(\mathbf{\mathcal{E}_p}^{(m)}\right),
    \label{eq:Aphi_mode}
\end{equation}
and 
\begin{equation}
   \mathbf{\mathcal{E}_p}^{(m)}= \mathcal{R}\left[\mathbf{u}^{(m)}\times \mathbf{B}^{(m)}|^{*}\right]-\mathcal{R}\left[\mathbf{u_p}^{(m)}\times \mathbf{B_p}^{(m)}|^{*}\right],
    \label{eq:EMF_mode_phi}
\end{equation}
and define
\begin{equation}
    \delta \mathcal{A}_\varphi^{\rm small} = \delta \mathcal{A}_\varphi - \sum\limits_{m=1}^{8}\delta\mathcal{A}_\varphi^{(m)}.
\end{equation}
In Fig.~\ref{fig:multiwav}b, we present the same decomposition as shown in Fig.~\ref{fig:multiwav}a, but now for the source terms of toroidal magnetic energy. These terms are calculated using Eqs.~(\ref{Eq:tor_ener}-\ref{eq:parseval_phi}-\ref{eq:EMF_mode_phi}), and all of them are subject to vertical averaging within the disk. Temporal averaging is performed between $5000,,r_g/c$ and $8000,,r_g/c$, and the values are normalized by the averaged axisymmetric toroidal magnetic energy, $\frac{1}{2}\mean{B_\varphi}^2$.

\begin{figure}
	\includegraphics[width=0.95\columnwidth]{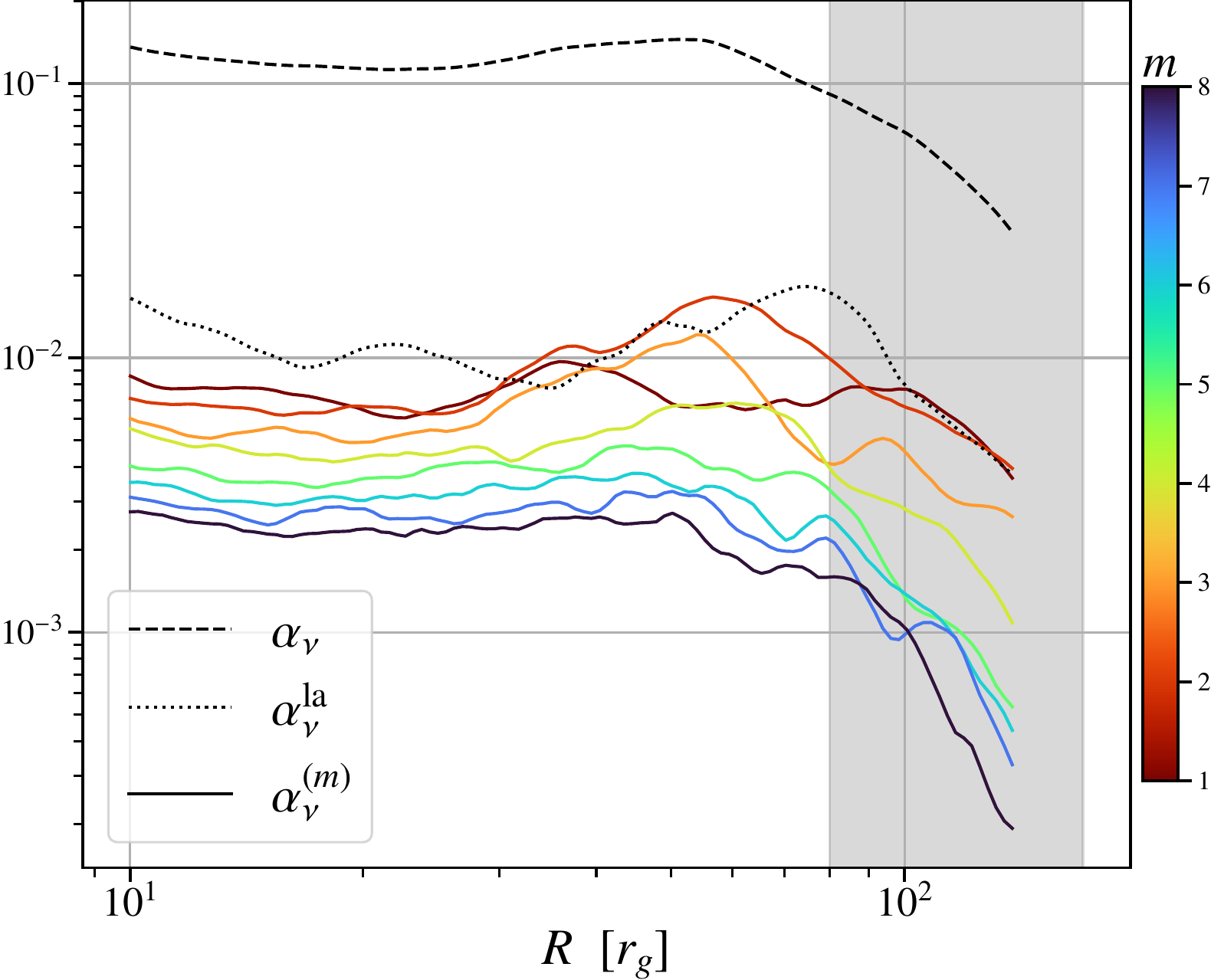}
    \caption{ Vertically and temporally averaged of the different component of the magnetic stresses computed from Eqs.~(\ref{eq:alpha_bet},\ref{eq:parseval_torque},\ref{eq:torque_m}). The temporal average is performed between $5000r_g/c$ and $8000r_g/c$ and the division between pressure and stress tensor is performed after both averages. The colored solid lines show the first eight non-axisymetric $m$-modes, $\alpha_nu^{(m)}$. The dotted line shows the large-scale axisymmetric transport of angular momentum by $-\mean{B_r}\mean{B_\varphi}$. The dashed line shows the small-scale turbulent transport of all non axisymmetric $m$-modes, $ \alpha_\nu=\sum\limits_{m=1}^\infty\alpha_\nu^{(m)}$ }
    \label{fig:m-torques}
\end{figure}

In line with Section \ref{sec:main_dinam}, we demonstrate that the sole term responsible for generating axisymmetric toroidal magnetic energy is $\mathcal{S}\varphi$, associated with the large-scale shear of the poloidal field. The large-scale advection term, $\mathcal{A}\varphi$, results in a loss of magnetic energy throughout the entire disk, corresponding to the vertical transport described in Section \ref{sec:main_dinam} and Section \ref{sec:quali}. All large-scale $m$-modes dissipate toroidal magnetic energy, in contrast to the decomposition of poloidal magnetic energy, here even the large-scale active modes ($m=1,2,3$) consume toroidal magnetic energy. This observation aligns with findings from shearing box simulations \citep{lesur_self-sustained_2008,riols_dissipative_2015,riols_magnetorotational_2017}. Once again, we observe that dissipation of the large scales is primarily driven by the more numerous small-scale $m$-modes, leading to the approximation that $\delta \mathcal{A}\varphi^{\rm small} \simeq \delta \mathcal{A}\varphi$.
\subsection{Angular momentum transport}
To show that the active modes transport angular momentum we decompose the Maxwell stress using Parseval's theorem
\begin{equation}
    \alpha_\nu = \sum\limits_{m=1}^\infty\alpha_\nu^{(m)},
    \label{eq:parseval_torque}
\end{equation}
where
\begin{equation}
   \alpha_\nu^{(m)}= -\frac{1}{4\pi\mean{P}}\mathcal{R}\left[B_r^{(m)}\times B_\varphi^{(m)}|^{*}\right].
    \label{eq:torque_m}
\end{equation}
Then, we define the large-scale laminar stress \citep{jacquemin-ide_magnetic_2021,manikantan_winds_2023}
\begin{equation}
    \alpha_{\nu}^{\rm la} = -\frac{1}{4\pi\mean{P}}\mean{B_r}\mean{B_\varphi}.
\end{equation}

In Figure \ref{fig:m-torques}, we present the $m$-mode decomposition of the magnetic angular momentum stresses using Eqs.~(\ref{Eq:tor_ener}, \ref{eq:parseval_phi}, \ref{eq:EMF_mode_phi}). All terms are subjected to vertical averaging within the disk and temporal averaging between $5000\,\,r_g/c$ and $8000\,\,r_g/c$; the division between pressure and stress tensor is performed after both averages. 
The active modes ($m=1,2,3$) contribute to angular momentum transport, but they are not the dominant stress term. The turbulent stress is primarily driven by the more numerous small-scale modes, this observation aligns with the contribution of the latter to turbulent dissipation.

\begin{figure*}
	\includegraphics[width=0.85\textwidth]{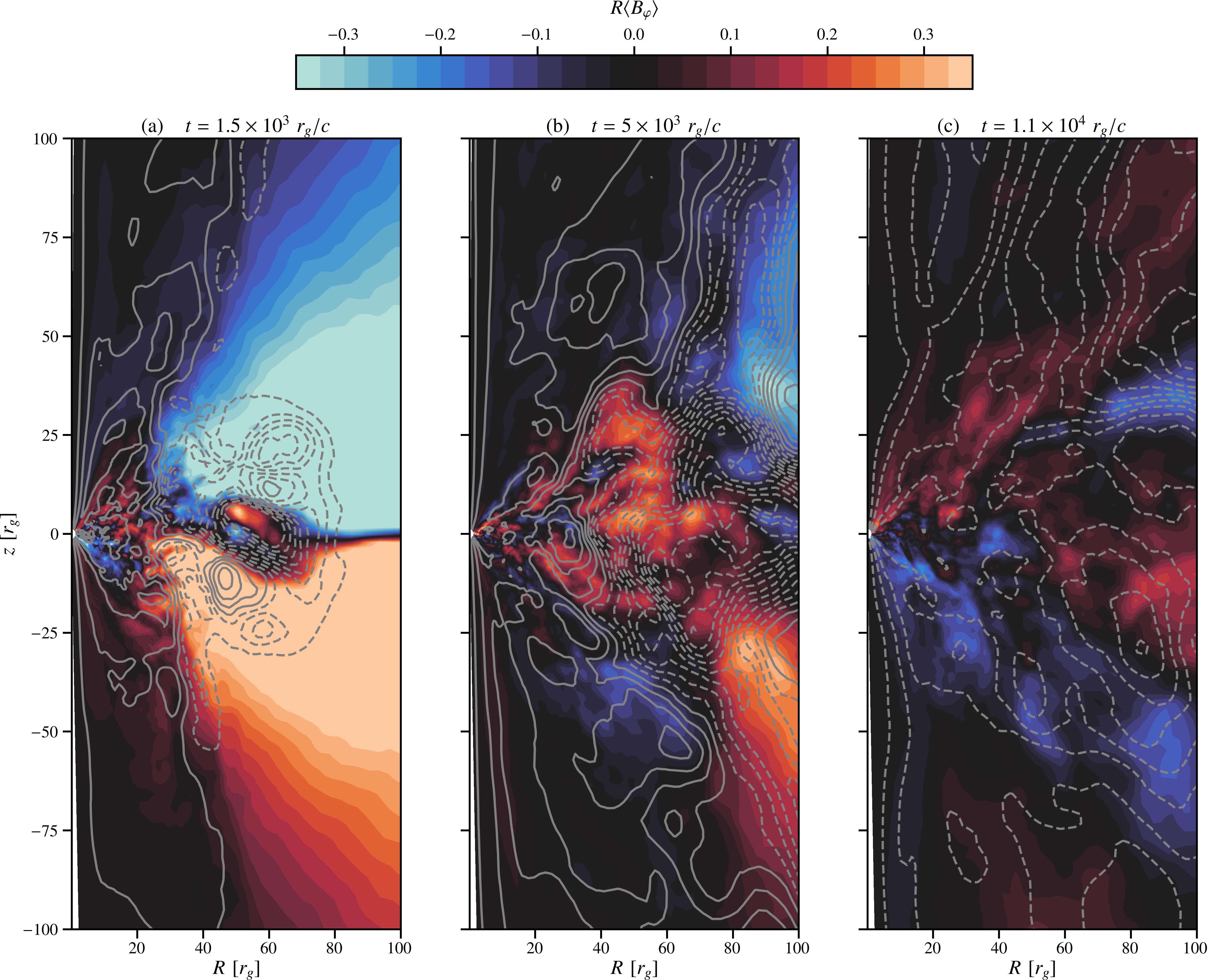}
    \caption{Snapshots of the axisymmetrized toroidal magnetic field, $R\mean{B_\varphi}$ in color as a function of $R$ and $z$. We also show the poloidal magnetic field lines through the poloidal magnetic flux (Eq.(\ref{Eq:pol_flux}), dashed lines show negative polarity and solid lines show positive polarity. Magnetic field structures become larger and larger with time. Notice that in this case the initial condition, visible at large radii in panel (a) is anti-symmetric with respect to the disk mid-plane.}
    \label{fig:aphi12}
\end{figure*}
\section{Simulation with an anti-symmetric toroidal field}\label{A:sim2}
In this appendix we describe the simulation initialized with an anti-symmetric, with respect to the disk midplane, toroidal magnetic field. This simulation shows identical features to the one described in the main body of the manuscript.

Figure \ref{fig:aphi12} show the axisymmetrized poloidal magnetic field lines and the axisymmetric toroidal magnetic field. We observe very similar evolution to that presented in Fig.~\ref{fig:aphi1}, even though the initial condition is very different: (1) large-scale magnetic fields are generated and vertically shed (see Fig.~\ref{fig:aphi12}a). (2) With time the size of the largest magnetic field structure increases, we notice advection of the larger field from the outer radii into the inner regions (Fig.~\ref{fig:aphi12}b,c).

Figure \ref{fig:aphi12} illustrates the axisymmetrized poloidal magnetic field lines alongside the axisymmetric toroidal magnetic field. Despite the contrasting initial condition, we observe a very similar evolution to that depicted in Fig.~\ref{fig:aphi1}: (1) The generation of large-scale magnetic fields, which are vertically expelled through ballooning instability \citep[][ see also Fig.~\ref{fig:aphi12}a]{lynden-bell_why_2003}. (2) Over time, an increase in the size of the magnetic field structures, with noticeable advection of the larger field from outer radii into the inner regions (see Fig.~\ref{fig:aphi12}b-c).

\begin{figure}
	\includegraphics[width=0.8\columnwidth]{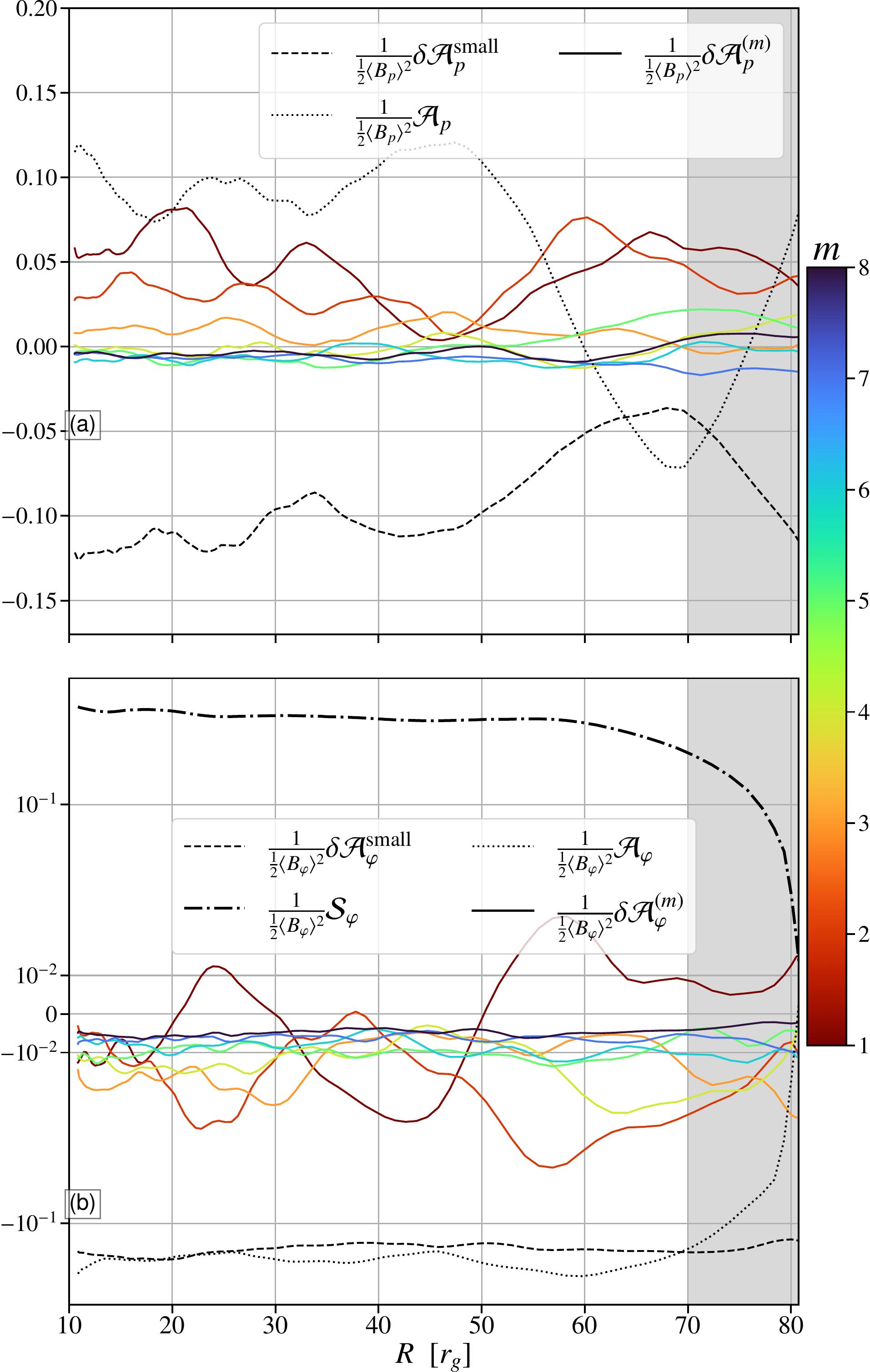}
    \caption{ Vertically and temporally averaged poloidal (a)  (Eq. \ref{Eq:pol_energ}), and toroidal (Eq.~\ref{Eq:tor_ener}) (b) energy equation normalized to the local Keplerian frequency. The temporal average is performed between $4000r_g/c$ and $7000r_g/c$. The different terms are divided by the vertical and temporal average of $\mean{B_p}^2$ (a) or $\mean{B_\varphi}$ (b). (a) The colored solid lines show the first eight non-axisymetric $m$-modes, $\delta\mathcal{A}_p^{(m)}$ (a) or $\delta\mathcal{A}_\varphi^{(m)}$ (b). The dotted line shows the large-scale axisymmetric transport of poloidal (a) or toroidal (b) field. The dashed line shows the small-scale turbulent transport without the first eight non-axisymetric modes, $ \delta \mathcal{A}_p^{\rm small}$ (a) or $ \delta \mathcal{A}_\varphi^{\rm small}$. The dotted-dashed line shows the shear of poloidal into toroidal field, $\mathcal{S}_\varphi$. }
    \label{fig:multiwav2}
\end{figure}

Figure \ref{fig:multiwav2}a shows the source terms for poloidal magnetic energy for the simulation initiated with an anti-symmetric toroidal field, mirroring the content of Fig.~\ref{fig:multiwav}a sor the symmetric case. The trends observed in the generation of poloidal magnetic field source terms align with the descriptions in Section \ref{sec:main_dinam} and Appendix \ref{A:multiwaves}.

Specifically, the findings reveal that in this case too only the largest non-axisymmetric modes ($m=1,2,3$) generate poloidal magnetic fields. Large-scale advection, $\mathcal{A}_p$, facilitates the transport of flux from outer regions, while the smaller non-axisymmetric modes ($m>3$) contribute to the dissipation of the poloidal magnetic field. Moreover, the amplitudes of these terms exhibit striking similarity with the symmetric case, yielding an identical regeneration timescale of approximately $t_{\rm g} = {\frac{1}{2}\mean{B_p}^2}/{\sum\limits_{m=1}^{3}\mathcal{A}_p^{m}}\simeq 10\Omega_K^{-1}$.

The patterns and amplitudes are also consistent for toroidal magnetic energy generation, as shown in Fig.~\ref{fig:multiwav2}b. Similarly to the poloidal magnetic energy case, the observations remain unchanged: all non-axisymmetric modes contribute to magnetic energy dissipation, magnetic energy generation primarily results from the shear of the poloidal field, and the toroidal magnetic field is shed vertically through the advective term.

We conclude that the dynamo mechanism is independent of the global symmetry of initial conditions.

\section{Time and lag diagram for correlations}\label{A:timelag}
\begin{figure*}
	\includegraphics[width=0.7\textwidth]{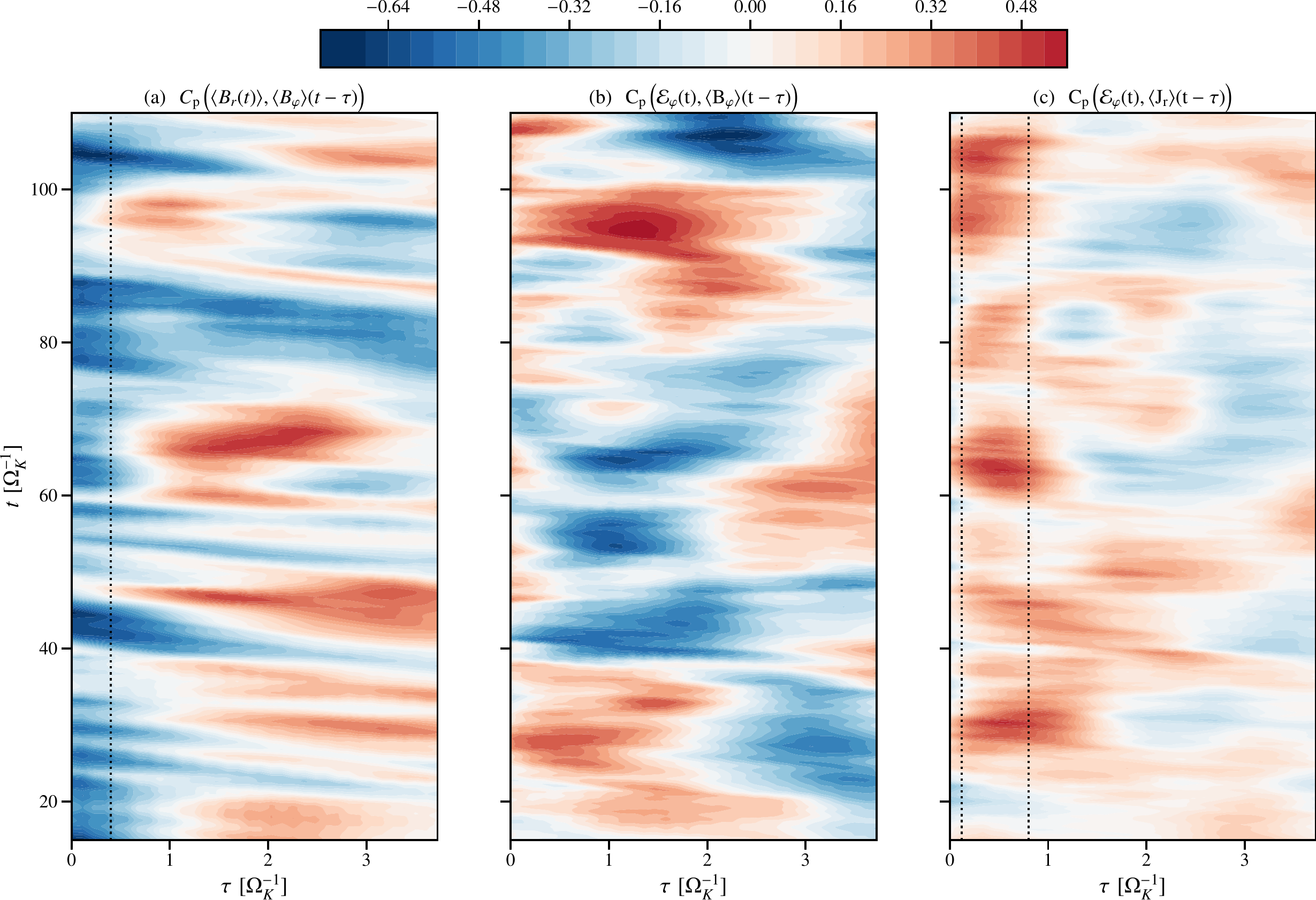}
    \caption{Vertically averaged lag-time diagram of the Pearson correlations at $r=20r_g$. The averages are computed between $t=20\Omega_K^{-1}(r=20)$ and $t=100\Omega_K^{-1}(r=20)$, and between $\theta_1 = \pi/2-\arctan(\frac{h}{r})$ and $\theta_2 = \pi/2-\arctan(0.1\frac{h}{r})$. The correlations are shown for: (a)the $\Omega$-effect which is used as a baseline and both models, (b) Eqs.~(\ref{eq:alpha_dyn}) and (c) (\ref{eq:shear_cur}).  Notice that on short time scales both models (b,c) show strong correlations, but the shear current effect (c) shows a more consistent correlation for $\tau=0.5\Omega_k^{-1}$.}
    \label{fig:correl_timelag_appen}
\end{figure*}
\bsp	
\label{lastpage}
\end{document}